\documentclass[preprint,12pt]{aastex}
\slugcomment{$$Id: ms.tex 1051 2009-08-27 00:27:01Z joishi $$}
\usepackage{amsmath}
\newcommand{\ud}{\mathrm{d}}

\begin{document}
\title{On Hydrodynamic Motions in Dead Zones}
\author{Jeffrey S. Oishi}
\affil{Department of Astronomy, 601 Campbell Hall, University of
  California at Berkeley, Berkeley, CA, 94720-3411}
\email{jsoishi@astro.berkeley.edu}
\and
\author{Mordecai-Mark Mac Low}
\affil{Department of Astrophysics, American Museum of Natural History,
  79th Street at Central Park West, New York, NY 10024-5192} 
\email{mordecai@amnh.org}

\begin{abstract}
  We investigate fluid motions near the midplane of vertically
  stratified accretion disks with highly resistive midplanes. In such
  disks, the magnetorotational instability drives turbulence in thin
  layers surrounding a resistive, stable dead zone. The turbulent
  layers in turn drive motions in the dead zone.  We examine the
  properties of these motions using three-dimensional, stratified,
  local, shearing-box, non-ideal, magnetohydrodynamical
  simulations. Although the turbulence in the active zones provides a
  source of vorticity to the midplane, no evidence for coherent
  vortices is found in our simulations. It appears that this is
  because of strong vertical oscillations in the dead zone. By
  analyzing time series of azimuthally-averaged flow quantities, we
  identify an axisymmetric wave mode particular to models with dead
  zones. This mode is reduced in amplitude, but not suppressed
  entirely, by changing the equation of state from isothermal to
  ideal. These waves are too low-frequency to affect sedimentation of
  dust to the midplane, but may have significance for the
  gravitational stability of the resulting midplane dust layers.
\end{abstract}


\section{Introduction}
A long standing problem in star formation concerns the accretion of
high-angular momentum material in disks around protostars. The
rediscovery of the magnetorotational instability (MRI) in an
astrophysical context, along with the discovery of the favorable
transport properties of the turbulence that results from it
\citep{HawleyGammieBalbus1995} led to the widespread belief that a
solution to this puzzle had been found. However, the instability
requires a critical coupling strength between field and fluid
\citep{BlaesBalbus1994,Jin1996}. Noting the high densities and low
temperatures likely at the midplanes of protoplanetary disks,
\citet{G96} developed a model including a weakly-coupled, MRI-stable
``dead zone'' at the midplane of such disks. Since this prediction,
considerable effort has gone into elucidating the conditions under
which such a model might exist. Most of this work has centered on
chemical studies of the conditions under which such a dead zone might
form and what its radial and vertical extent might be. Recently, a
number of numerical studies, pioneered by \citet{FlemingStone2003},
have begun to explore the dynamical consequences of the dead zone
model \citet{FromangPapaloizou2006}. \citet{TurnerSanoDziourkevitch2007}
have developed an interesting coupled chemical and
magnetohydrodynamical (MHD) model in which the dead zone disappears
because of vertical mixing of ionized gas to the midplane. In their
picture, the dead zone is absent in the sense that large scale
magnetic fields with positive Maxwell stress, $\langle -B_r B_\phi
\rangle$, form in the midplane and transport significant amounts of
angular momentum even though the midplane is still MRI stable.

Aside from acting as a conduit for material flowing onto the
protostar via accretion, protoplanetary disk midplanes are also 
the sites of planet formation. One of the outstanding
issues in planet formation is the collection of micron-sized dust into
planetesimals, rocky objects of order $1 \mbox{ km}$ in size. In this
context, turbulent fluid motions more general than the correlated
fluctuations that drive accretion become very important. This is
because drag forces couple the dust and gas.  Fully MRI-active
simulations, \citet{JohansenKlahr2005} find preferential concentration
of dust in turbulent over-pressures. However, these same turbulent
motions increase the velocity dispersion of the dust grains, possibly
leading to destructive collisions. The dust-gas coupling can drive
strong instability and clumping
\citep{YoudinJohansen2007,JohansenYoudin2007}, which, in combination with
MRI turbulence, allows direct formation of planetesimals by
gravitational instability \citep{Johansen+etal2007}. Together, these
results suggest that turbulence plays a complex role in planetesimal
formation. As a first step to considering more complex dust-gas models
including dead zones, it is critical to characterize the fluid motions
in the MRI-stable midplane.

Here, we address two specific questions about the fluid motions in
dead zones: do coherent, two-dimensional (2D) vortices form in dead zones?
And, what are the wave-like motions seen in our previously published models of dead
zone dynamics \citep[e.g.][]{OishiMacLowMenou2007}?  Vortices have been studied both
for their ability to drive purely hydrodynamic accretion
\citet{UmurhanRegev2004} and their ability to trap dust particles and
thus act as sites of enhanced planetesimal formation
\citep{Bracco+etal1999,Johansen+etal2004}.  \citet[][hereafter
JG05]{JG05b} studied vortex dynamics in a 2D,
compressible, shearing sheet model and concluded that random vorticity
perturbations do form coherent large-scale vortices, but that energy
in such structures decays away as $t^{-0.5}$. They also suggested that
turbulent overshoot from the active layers might provide a vorticity
source at the midplane. Using three-dimensional (3D), compressible but
unstratified simulations, \citet{Shen+etal2006} showed that small
amplitude vortices are unstable to elliptical instabilities
\citep{Kerswell2002} and that finite-amplitude trailing waves are torn
apart by a Kelvin-Helmholtz like instability before they have a chance
to form coherent vortices. The anelastic simulations of
\citet{BarrancoMarcus2005} show large, coherent vortices forming at
1--3~$H$ from the midplane, where $H$ is the disk scale height. This
unusual and quite unexpected result suggests that stratification plays
an important role in disk vortex dynamics. Because our time-steps are
Courant-limited by the Alfv\'en velocity, we restrict our domain to
$|z| < 2 H$ to avoid low-density regions with extremely high Alfv\'en
speeds. Nonetheless, we study the effect of stratification on the flow
inside and outside of the dead zone with the goal of understanding if
MRI turbulence in active layers can sustain vortices.

In section~\ref{s:method} we briefly describe our numerical techniques
and how we tested them, while section~\ref{s:results} contains the
results of our 3D simulations, and section~\ref{s:vortex} reviews
aspects of vortex dynamics and presents relevant 2D
simulations. Finally, we discuss our results and present conclusions
in section~\ref{s:discussion}.

\section{Numerical Method}
\label{s:method}
\subsection{Model Description}
We use the Pencil Code \citep{BrandenburgDobler2002} to solve the
resistive MHD equations in the vertically stratified, shearing box
limit \citep{HawleyGammieBalbus1995}. 

All models are 3D, isothermal, non-ideal MHD, except for one ideal MHD
control run, and one non-ideal, non-isothermal run (see
section~\ref{s:ideal_gas}). They are run on a domain of $1 H \times 4
H \times 4 H$, where $H$ is the disk scale height. The numerical
method is stabilized with sixth-order hyperviscosity,
hyperresistivity, and a hyperdiffusive operator on the density as
described in \citet{OishiMacLowMenou2007}. All the models are
initialized with a MRI unstable zero-net flux initial magnetic field
given by $B_z(x) = B_0 sin(2 \pi x /L_x)$ with initial plasma $\beta =
2 \mu_0 P_g /B_0^2 = 400$. The domain has periodic boundary
conditions in $y$ and $z$, and shearing periodic boundaries in $x$.

The non-ideal models use one of two $z$-dependent resistivity
profiles. One is motivated by balancing cosmic ray-ionization to a
critical depth in column density given by
\citet{FlemingStone2003}. These models were previously described in
\citet{OishiMacLowMenou2007} and are hereafter referred to as ``FS
runs''. These models were run with cubical zones of size $32$ and $64$
zones per scale height. Unless otherwise noted, we use a resolution of
$64$ zones per scale height for the FS runs. The second profile,
described below, uses hyperbolic tangent functions and models using it
were run with $32$ zones per scale height. 

\subsection{Hyperbolic Tangent Resistivity Profile}
This alternate profile offers a sharper transition between active and
dead zones than the FS profile, and allows the thickness of the dead zone to be controlled
independently from its depth. Though simpler, it is less physically motivated than the FS
profile. The profile is given by
\begin{equation} \label{eq:tanh}
  \eta(z) = \frac{\eta_0}{2} \left( \tanh\left(\frac{z+z_0}{\Delta z}\right) - \tanh\left(\frac{z-z_0}{\Delta z}\right) \right)
\end{equation}
where $\eta_0$ is the midplane resistivity, $z_0$ is the transition
height and $\Delta z$ is the width of the transition region, set in
this work to $\Delta z = 5 dx$, where grid zones have size $dx$. We
set $\eta_0 = 1.67 \times 10^{-5}$, corresponding to magnetic Reynolds
number $Re_M = c_s^2/(\eta_{mid} \Omega) = 30$  at $z < |z_0|$. This gives all our $\tanh$ models
the same value of $Re_M$ in the dead zone as our fiducial FS run has
at $z = 0$.

\begin{figure}[htbp]
  \includegraphics[scale=0.55]{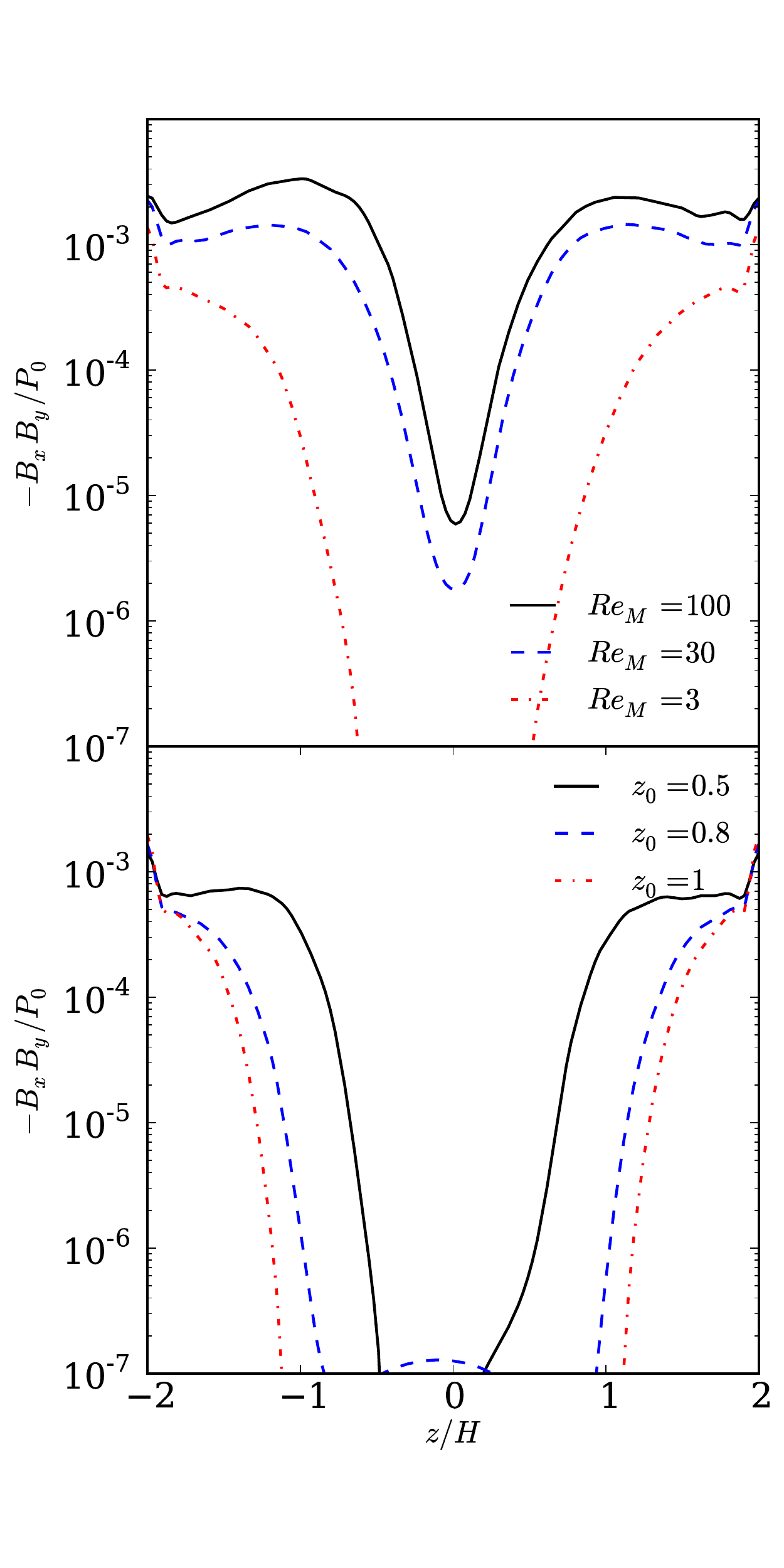}
  \caption{
    \label{f:maxw_z_32}
    Maxwell stress as a function of height $z$ for three different
    values of magnetic Reynolds number using the FS profile, and three
    different transition heights $z_0$ using the $\tanh$ resistivity
    profile (Eq.\ \ref{eq:tanh}). At very high $\eta$ values, the
    stress is occasionally positive and therefore absent on this plot.
  }
\end{figure}

The Maxwell stresses as a function of height for both sets of profiles
are given in Figure~\ref{f:maxw_z_32}. The figure emphasizes the
main utility of the $\tanh$ profile: because the dead zone resistivity
is independent of its size, we can study larger dead zones at a given
resolution.\placefigure{f:maxw_z_32}
\subsection{Tests}

In the numerical study of non-axisymmetric perturbations in disks, a
series of test problems are becoming standard. The vortical and
non-vortical shearing waves have closed analytic solutions (in the
former case, a WKB approximation) derived by \citet{JG05a}. These
waves have been used by JG05 to test a code derived from ZEUS
\citep{StoneNorman1992} and the higher-order Godunov code ATHENA
\citep{Shen+etal2006}. Aside from providing an analytic test solution
for comparison, the incompressible wave is particularly useful for
determining the amount of aliasing present in the numerical scheme. As
a wave swings from leading to trailing, it wraps toward
axisymmetry. However, in doing so the number of zones resolving the
wave must drop. When the number of zones resolving each wavelength
drops below some threshold, the code may unphysically transfer
(``alias'') power from the trailing wave into a new leading wave that
will again swing amplify. This danger is highlighted by both
\citet{UmurhanRegev2004} and \citet{Shen+etal2006}, as such power may
spuriously drive coherent vortices and angular momentum transport.

The Pencil Code uses spatially-centered finite differences and
Runga-Kutta timestepping, meaning that its discretization lacks any
formal algorithmic viscosity. Thus, it should perform well on this
test. Ultimately, however, this property works against the code:
without an explicit viscous term, it does not dissipate energy that moves
to smaller scales, where an unphysical buildup occurs. In this sense,
the shearing waves mimic the effect of a turbulent cascade: as time
progresses, $k_x(t)$ decreases and the waves become less and less
resolved. This is not a turbulent cascade of course, but the code is
nonetheless forced to resolve smaller and smaller structures, leading
eventually to a crash. To avoid this in these tests, as in our actual
runs, we use a sixth-order hyperviscosity scaled by $\nu_6$ for
stability, setting the hyperviscous Reynolds number at the grid scale
$Re_6 = u dx^5/\nu_6 \simeq 1$, where $u$ is the maximum velocity on
the grid.

\begin{figure}[htbp]
  \includegraphics[scale=0.8]{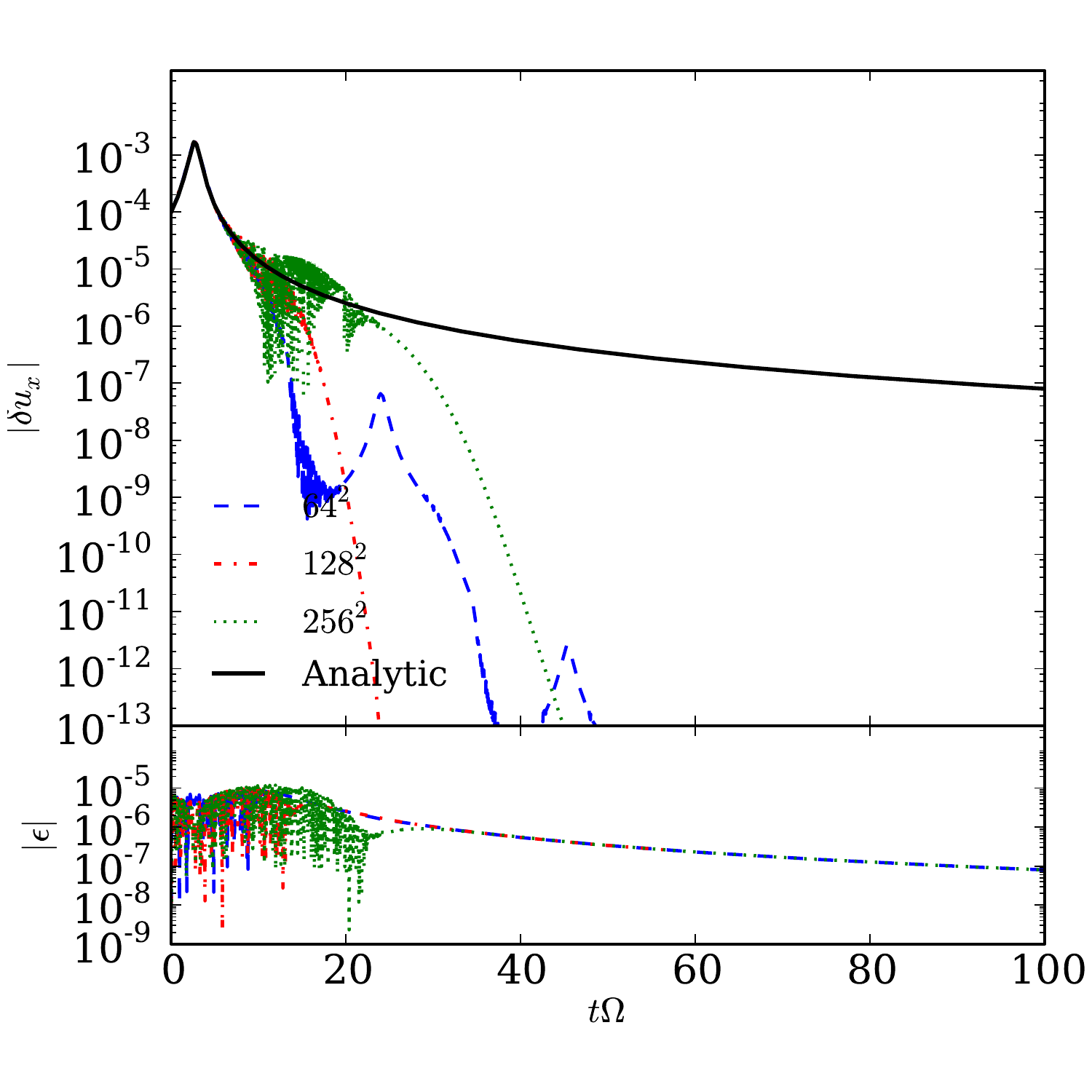}
  \caption{
    \label{f:incomp_shwave_t_res}
    Time evolution of the velocity perturbation amplitude $\delta u_x$
    of an incompressible, shearing wave (upper panel) using physical
    viscosity for resolutions from $64^2$ to $128^2$. The analytic
    solution is given in black. Aliasing is present in the $64^2$
    solutions but injects only trivial amounts of energy. The lower
    panel shows absolute value of the error, $|\epsilon| = |\delta u_x
    - \delta u_x^{analytic}|$. Note that this error smoothly
    asymptotes to the analytic solution as the code hyperviscously
    damps the wave faster than aliasing can inject spurious energy.}
\end{figure}

Figure~\ref{f:incomp_shwave_t_res} shows the incompressible shearing
wave for runs with resolution $N = 64, 128, 256$, demonstrating the
ability of the Pencil
code to accurately reproduce the analytic solution. The aliasing time
in terms of the orbital frequency is \citep{Shen+etal2006}
\begin{equation}
t_{alias} \Omega_0 = \frac{n}{q} \left(\frac{N_x}{n_y} -\frac{k_{x0}}{k_y}\right),
\end{equation}
where $n$ is an integer, $q = \mathrm{d}\ln \Omega_0/\mathrm{d}\ln r$
is the shear parameter, and $n_y = k_y L_y/2 \pi$ is the dimensionless
azimuthal wavenumber of the wave. The oscillations around $t=20$
appear to be small amplitude sound waves present in our compressible
solution but not in the incompressible analytic solution.

\begin{figure}[htbp]
  \includegraphics[scale=0.8]{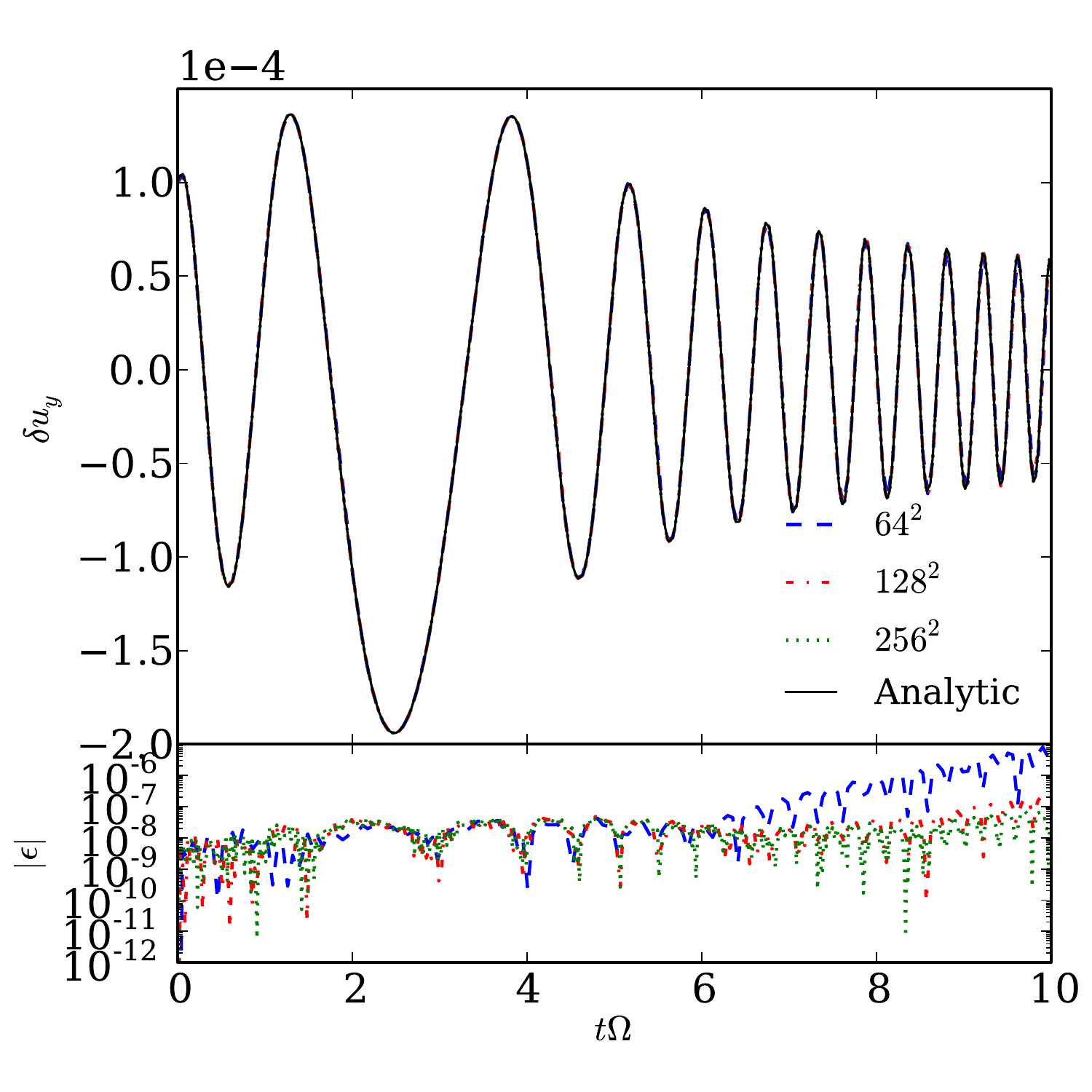}
  \caption{
    \label{f:comp_shwave_t_res}
    Time evolution of the amplitude $\delta u_y$ of a compressible,
    shearing wave using hyperviscosity for resolutions from $32^2$ to
    $128^2$. The analytic solution is given in black.   The lower panel
    shows the absolute value of the error, $|\epsilon| = |\delta u_y -
    \delta u_y^{analytic}|$, between the Pencil Code solution and a
    numerically integrated solution to the exact parabolic cylinder
    equation describing the wave.  }
\end{figure}

In Figure~\ref{f:incomp_shwave_t_res}, the $N = 64$, $128$, and $256$
models run to $t \Omega_0 = 100$, where $\Omega_0$ is the local
orbital frequency.  This is well beyond the expected aliasing time
$t_{alias} \Omega_0 \simeq 45.3 n$ for $N_x = 128$ and $90.6 n$ for
$N_x = 256$. The figure makes an interesting point: aliasing does
indeed occur for $64^2$, but because the hyperviscosity is a strongly
non-linear function of resolution, the spurious energy injected by
aliasing becomes trivial before the first $128^2$ aliasing time. Even
at $64^2$, each successive aliased wave has less power, suggesting
that sustained vortex activity is not likely to be powered by this
numerical effect for long times. There is no aliasing at all at
$128^2$ and $256^2$ resolutions, because the hyperviscosity damps the
signal to machine precision before the first aliasing time. We have
additionally performed a convergence test using resolutions of $64$,
$128$, and $256$. Using hyperviscosity, the average error over $t
\Omega < 10$ reaches a constant value at $128^2$ and does not further
converge. However, when we stabilize the algorithm using a physical
Laplacian viscosity instead of hyperviscosity, $\mathbf{f_\nu} = \nu
\nabla^2 \mathbf{u}$, the error converges at roughly $\sim N^{-1.6}$
where $N$ is the resolution.  We also consider the compressible
shearing wave. Figure~\ref{f:comp_shwave_t_res} demonstrates the
code's ability to resolve the wave with small error to late times with
moderate resolution.

\begin{figure}[htbp]
  \includegraphics[scale=0.8]{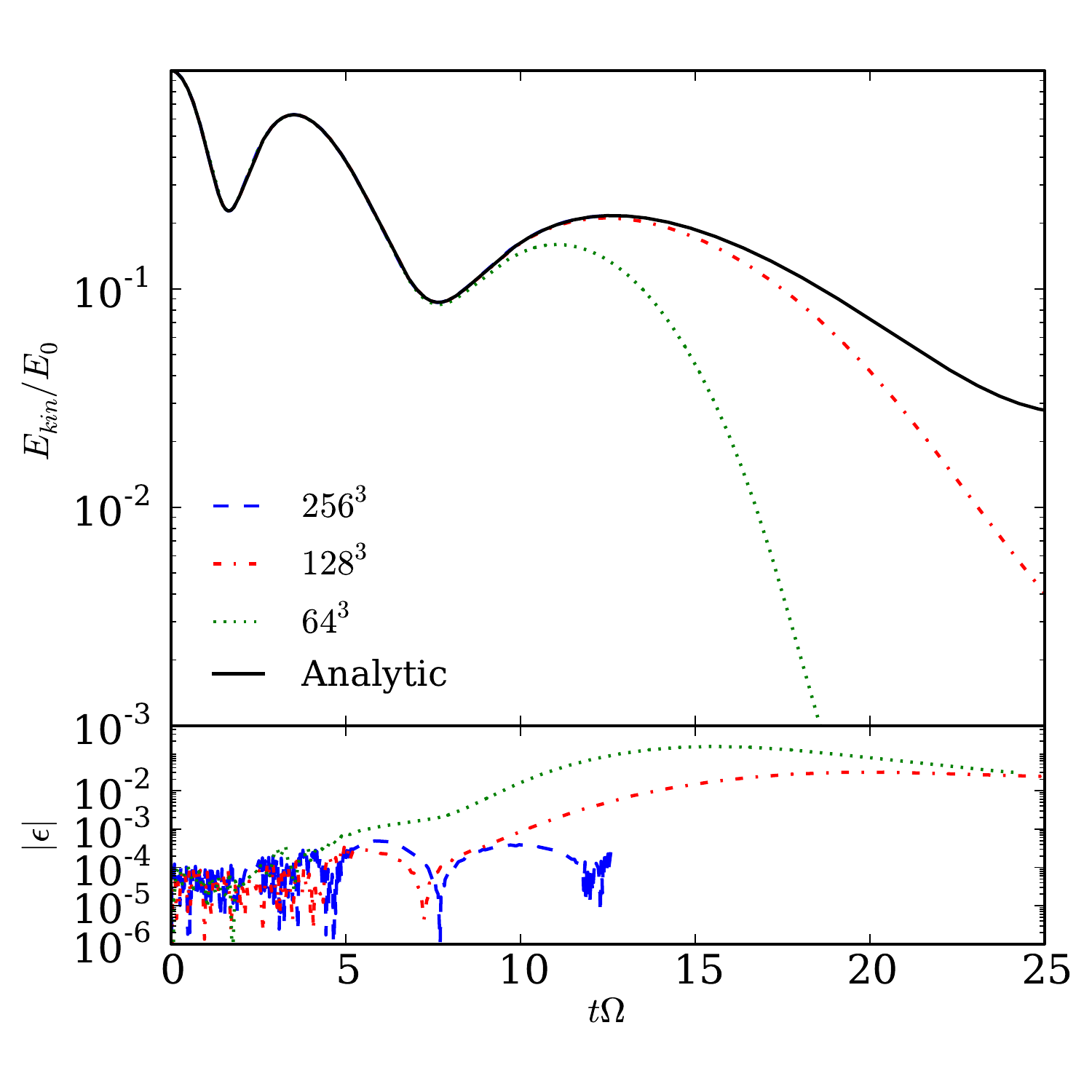}
  \caption{
    \label{f:bh_t_res}
    Time evolution of the kinetic energy of a 3D, non-linear,
    incompressible, shearing wave using hyperviscosity for resolutions
    from $64^3$ to $256^3$. A numerical integration of the exact wave
    differential equation is given in black and labeled ``Analytic''. The lower panel
    shows absolute value of error, $|\epsilon| = |E_{kin} - E_{kin}^{analytic}|$
  }
\end{figure}

Finally, we successfully tested against the 3D, hydrodynamic,
non-linear solution of the shearing box equations derived by
\citet{BalbusHawley2006}. Figure~\ref{f:bh_t_res} shows the kinetic
energy of the wave as function of time. In this test, as in the
others, the wavenumber drops, so that at larger times the error
necessarily grows.

\begin{figure}
  \plotone{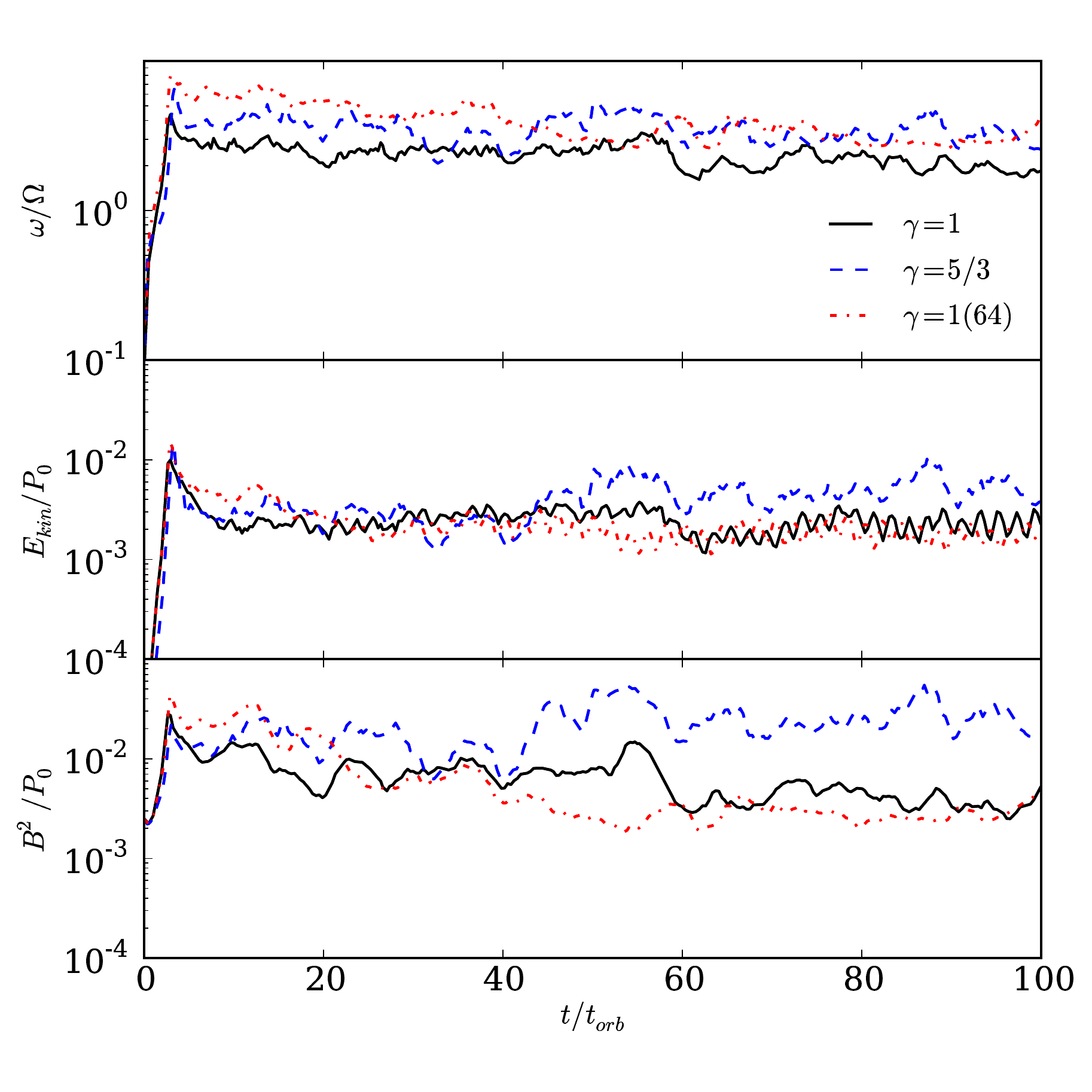}
  \caption{From top to bottom: volume averaged vorticity ($\omega$),
    kinetic, and magnetic energy comparing isothermal and ideal gas
    $Re_M = 30$ runs. Isothermal runs at $H = 64 dx$ and $H = 32 dx$
    are overplotted. The amount of vorticity increases with increasing
    resolution.}
  \label{f:en_adi_iso}
\end{figure}

The incompressive shear wave test gives us a strict metric for
aliasing: for the lowest resolution, we used $H = 64 dx$, and saw
trivial amounts of aliasing. Doubling the resolution to $H = 128 dx$,
we see no aliasing. Because the amount of vorticity \emph{increases}
in our MHD simulations when we increase the resolution from $H = 32 dx$
to $64 dx$ (see figure~\ref{f:en_adi_iso}), we are confident that any
sustained vorticity in our MHD simulations does not come from
numerical aliasing, and that the code accurately tracks any vortical
motions present.

\section{Results}
\label{s:results}
\begin{figure}
  \plotone{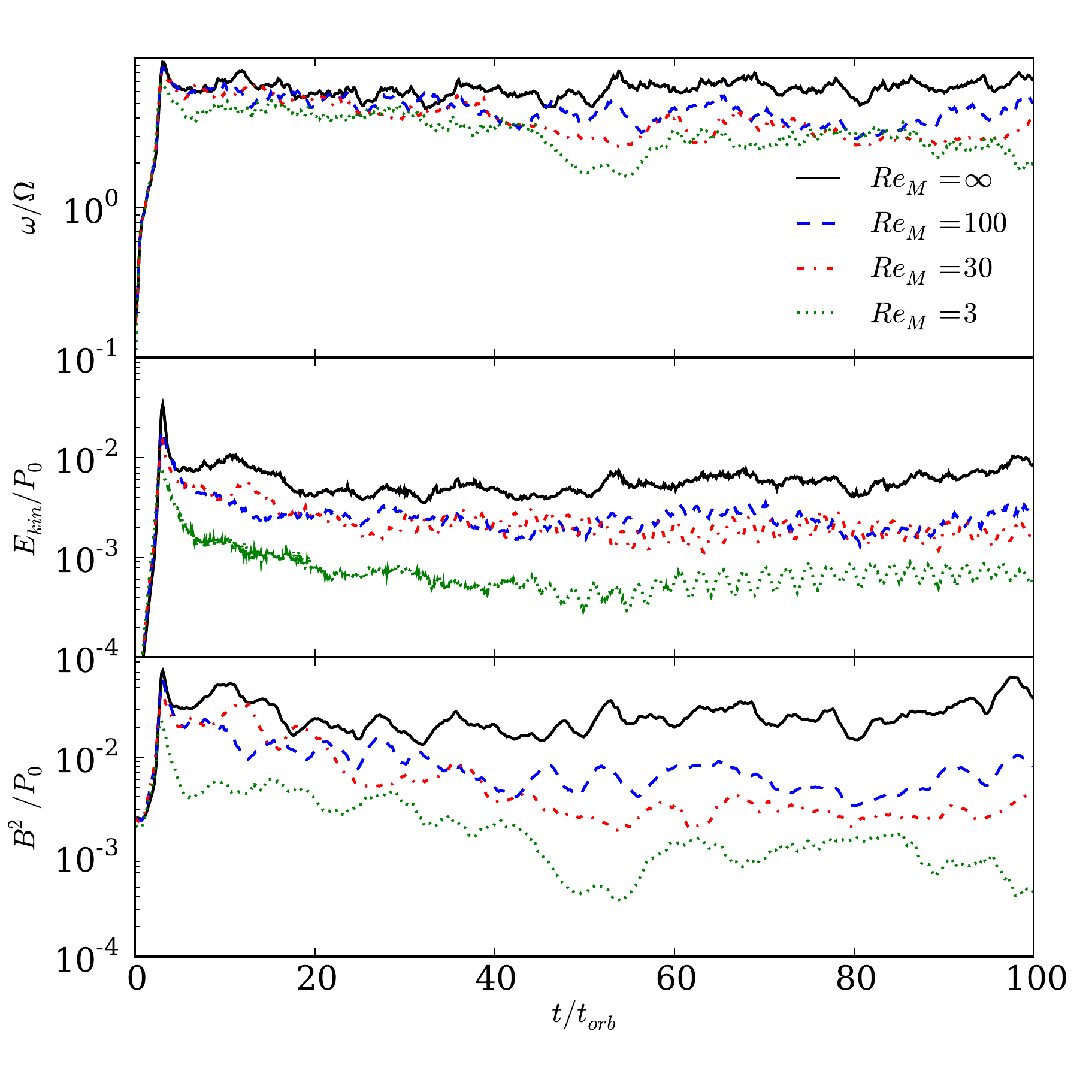}
  \caption{ Same as figure~\ref{f:en_adi_iso} for each $H = 64 dx$
    resolution FS run.}

    \label{f:M64_energy_alpha_vorticity}
\end{figure}

The excitation and saturation of the MRI occur by $t/t_{orb} = 5$ and
the dead zone is clearly defined by orbit $\sim 10$ (see section~3.3
for details). Here, we will restrict our analysis to times later than
$25$ orbits in order to avoid 
transients. Figure~\ref{f:M64_energy_alpha_vorticity} shows time
series of vorticity ($\mathbf{\omega = \nabla \times u}$), and kinetic
and magnetic energies. Although the saturated kinetic energy drops by
over an order of magnitude between the $Re_M = \infty$ and the $Re_M =
3$ models, the vorticity only drops by a factor of a few. This can be
explained by the presence of residual vorticity in the dead zone.

\subsection{Vorticity and Flow Dimensionality}
The MRI produces significant amounts of vorticity, much of which is
retained when a dead zone is introduced. Here, we investigate whether
or not this vorticity can coalesce into coherent vortices that
could be significant as a natural environment in which to trap dust
and thus accelerate protoplanet formation.

In a thin accretion disk, the vertical gravitational acceleration is given by
\begin{equation}
  \mathbf{g} = - \Omega^2 z \mathbf{\hat{z}},
\end{equation}
where $\Omega$ is the orbital frequency. In the local approximation,
we take $\Omega_0 = \Omega(R)$ where $R$ is the assumed central radius
of the shearing box. For an isothermal disk in hydrostatic
equilibrium, the Brunt-V\"ais\"al\"a frequency of buoyant vertical oscillation is
\begin{equation}
\label{e:brunt}
N^2 = -\frac{\mathbf{g}}{\rho} \frac{\partial \rho}{\partial z} = \frac{\Omega_0^2 z^2}{ H^2}.
\end{equation}
We can diagnose how stratified the flow is by comparing the vertical
component of the vorticity $\omega_z$ to the Brunt-V\"ais\"al\"a
frequency $N$ in a version of the internal Froude number, \citep[e.g.,][]{BarrancoMarcus2005}
\begin{equation} \mbox{Fr} =\omega_z / (2 N). \end{equation}
$N$ vanishes at
the midplane, and thus Fr formally diverges as $z \to 0$. Nonetheless,
Fr provides a useful diagnostic for the degree of stratification in
the flow: when Fr$ < 1$, internal gravity waves rapidly homogenize
vertical disturbances in the vortex, so the system is strongly
stratified, and acts effectively as a 2D flow in which energy cascades
to larger scales, similarly to the Earth's atmosphere. Thus, Fr is an
effective measure of the dimensionality of a vortex in the disk
plane. 

\begin{figure}[htbp]
  \plotone{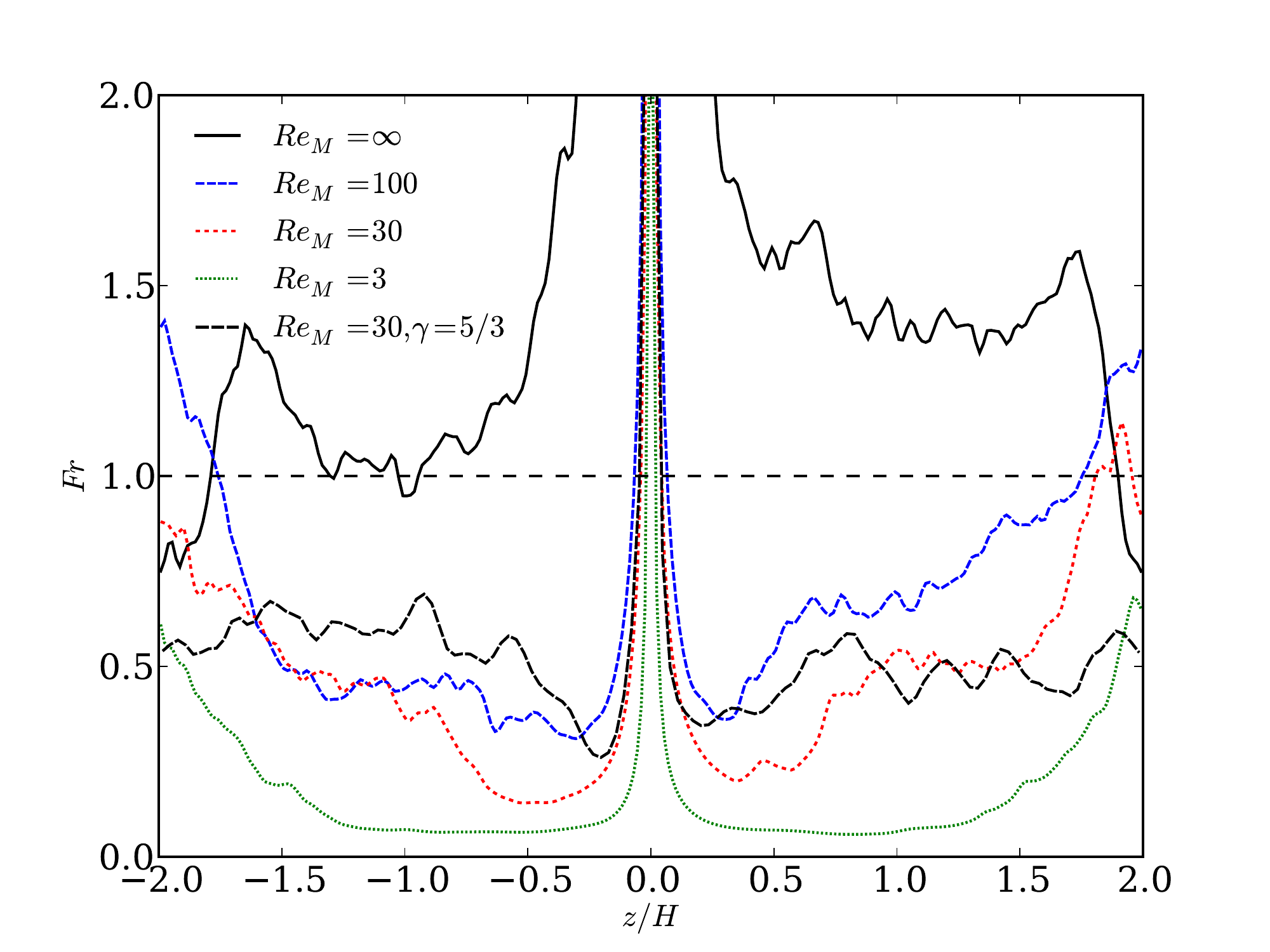}
  \caption{
    \label{f:Fr_vs_z}
    Internal Froude number versus height above the midplane for the FS
    runs.  At Fr$ \lesssim 1$, the flow becomes strongly stratified
    and effectively two-dimensional for vortices lying in the disk
    plane.}
\end{figure}

Previous results show rapid vortex growth in purely 2D systems
integrated over $z$ \citep{UmurhanRegev2004,JG05b}, and at $Fr < 1$ in
3D stratified systems
\citep{BarrancoMarcus2005}. Figure~\ref{f:Fr_vs_z} shows the vertical
behavior of $Fr$ in our FS models. All but one of our simulations are
purely isothermal, with $P = c_s^2 \rho$, so gravity waves themselves
are excluded. However, because the basic dynamical properties of the
MRI are not very sensitive to the equation of state
\citep{StoneHawleyGammieBalbus1996}, we do not expect that including
gravity waves will change the $Fr$ profile. We confirm this
expectation in figure~\ref{f:Fr_vs_z}, which also includes our control
$Re_M = 30$ run with a non-isothermal, $\gamma=5/3$ ideal gas equation
of state (see section~\ref{s:ideal_gas} for more details). The $Fr$
profile is roughly similar to the run with isothermal $Re_M = 30$.

In our ideal MHD run the flow is effectively 3D through almost the
entire domain---the MRI is a strong enough vorticity source to
overwhelm the homogenizing effect of vertical oscillations almost
everywhere.  On the other hand, Figure~\ref{f:Fr_vs_z} suggests that
the dead zone models have an effectively 2D flow at nearly all
heights, most particularly in the dead zone of the $Re_M = 3$
model. Near the midplane, $Fr$ diverges, as the stratification there
goes to zero, so we would expect the very near midplane region to act
as a 3D flow, consistent with the \citet{BarrancoMarcus2005} results
showing disappearance of their imposed vortices in that region.

\begin{figure}[htbp]
  \plotone{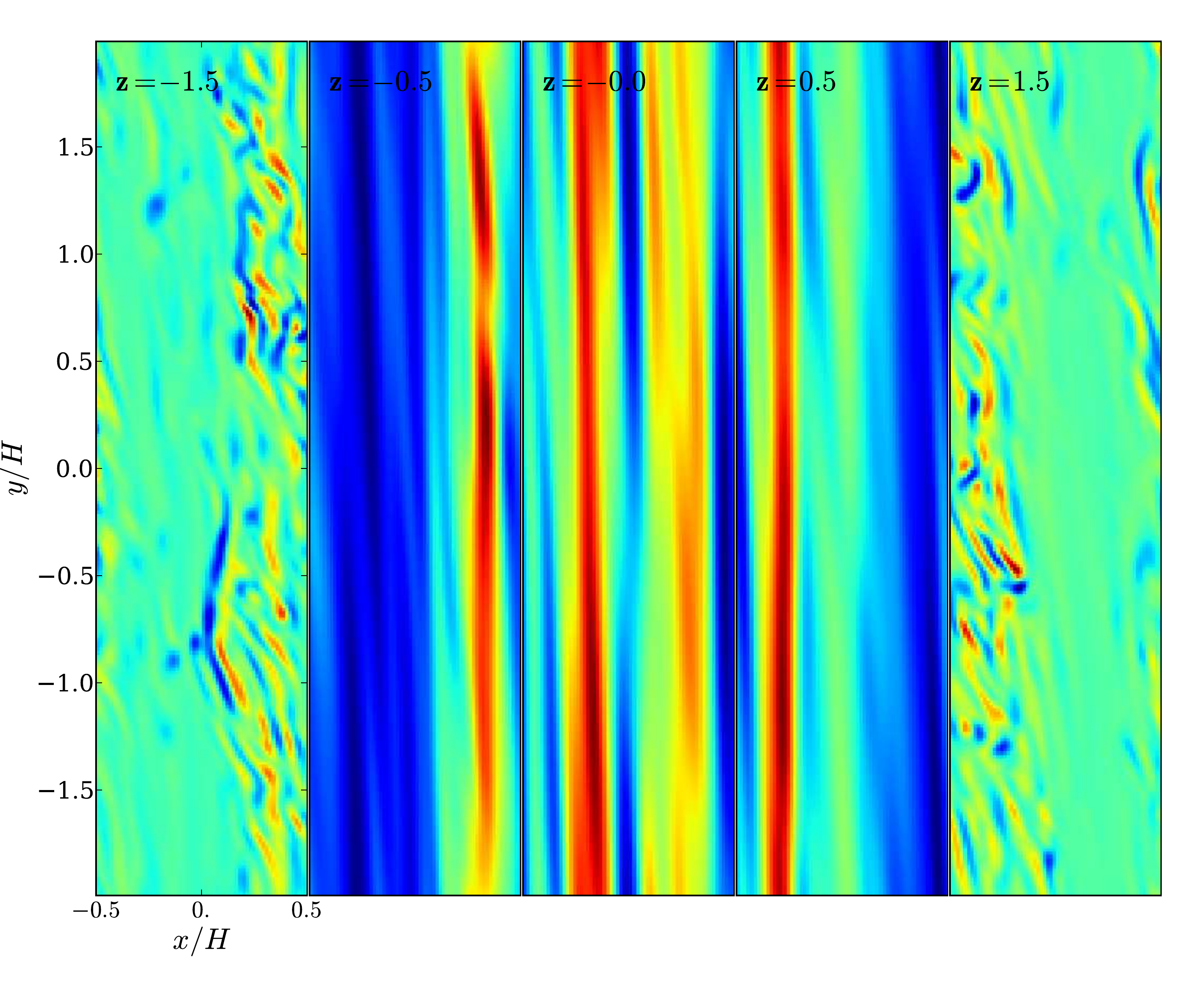}
  \caption{
    \label{f:img_vorticity_64_Re_3_geta}
    Vorticity in horizontal ($x$-$y$) planes as a function of height
    above and below the midplane for the $Re_M = 3$ run. The effectively
    2D nature of the flow does not appear to create regions in
    which coherent vortices can form.
  }
\end{figure}

However, Figure~\ref{f:img_vorticity_64_Re_3_geta} shows that no
vortices form in either the dead or the active zone.  In the dead
zone, the vorticity remains confined to elongated, nearly axisymmetric
bands.  In fact, no difference is seen in $\omega_z$ between heights
within the dead zone--only between dead and active zones. Thus, our
results combined with \citet{Shen+etal2006} and
\citet{BarrancoMarcus2005}, show that even in the presence of a
vorticity source such as active zones, stratification is necessary but
insufficient to produce vortices.

Table~\ref{t:enstrophy} shows the distribution of enstrophy among
$x$,$y$, and $z$ components for FS models and the ideal MRI run. A
strongly two-dimensional flow shows a preferred direction
\citep{BNST95}. The ideal MRI run and the active zones of the FS
models agree well with an isotropic distribution of vorticity. The
dead zones show a strong preference for the $y$ direction, indicating
a circulation in the $x-z$ plane, as indeed images bear out (see
below). The coherent anticyclonic vortices we set out to investigate
would have shown a strong $z$ component.

\begin{figure}
  \plotone{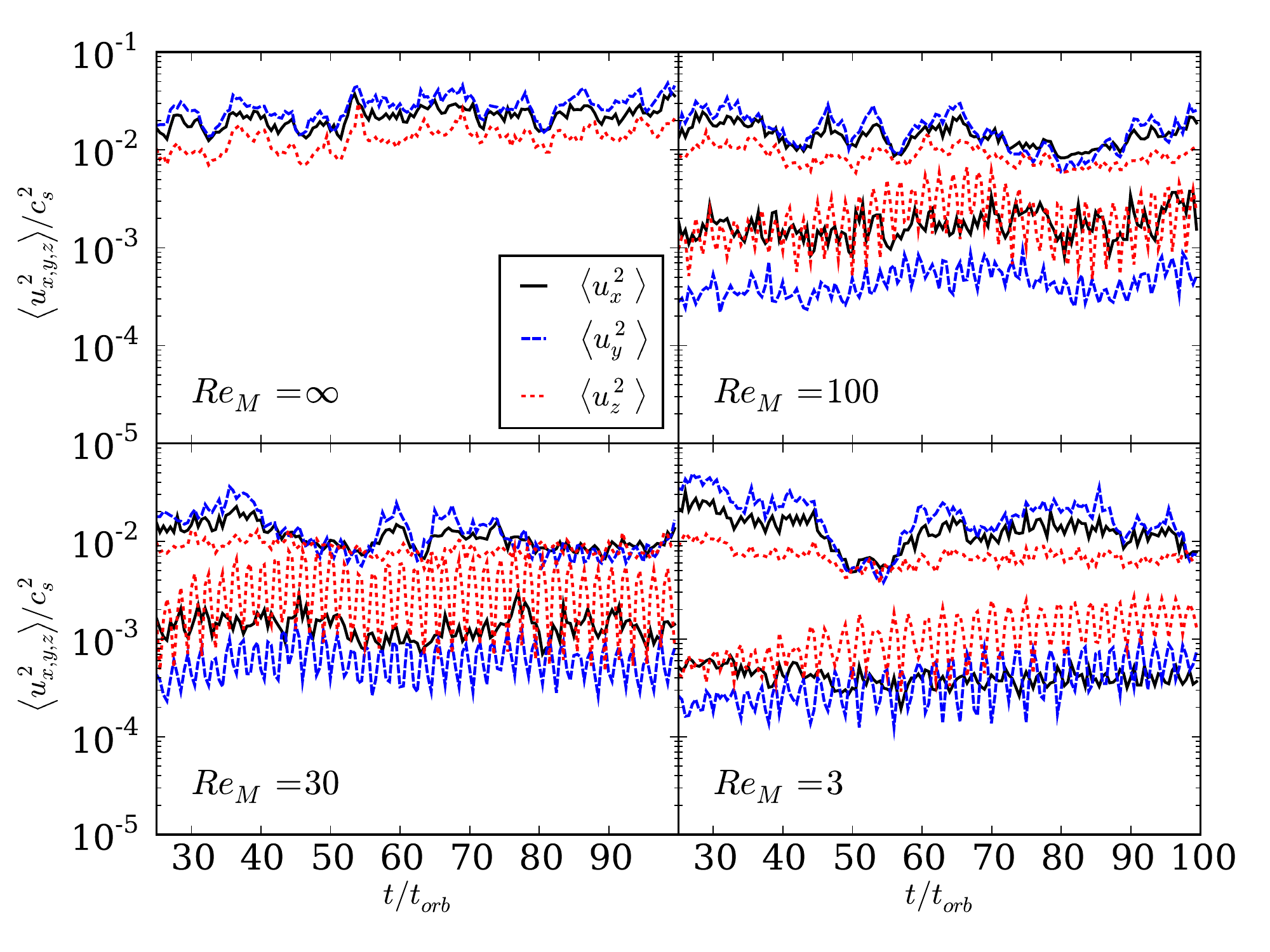}
  \caption{
    \label{f:M64_ux2_uy2_uz2_act_dead}
    Mean square velocities in each direction for the FS runs. In all
  cases with a dead zone, the upper three curves show the active zone
  and the lower three the dead zone velocities. Note the transition in
  dominant component from $y$ velocity to $z$ velocity from active to
  dead in all cases.
  }
\end{figure}

Figure~\ref{f:M64_ux2_uy2_uz2_act_dead} shows the
components of velocity dispersion for each of the FS runs as a
function of time, computed separately inside and outside of the dead
zone. The $z$ component of velocity dominates within the dead zone,
while outside of it, it is a factor of a few smaller than the
others. (Note that the oscillation in the dead zone velocities is real, but the
plot was made with relatively crude time resolution, $\delta t = 0.5
t_{orb}$ and some aliasing from higher frequency signals is certainly
present.) The dead zone motion is clearly not dominated by
horizontal flows.

\subsection{Morphology and velocity vectors}
\begin{figure}
  \plotone{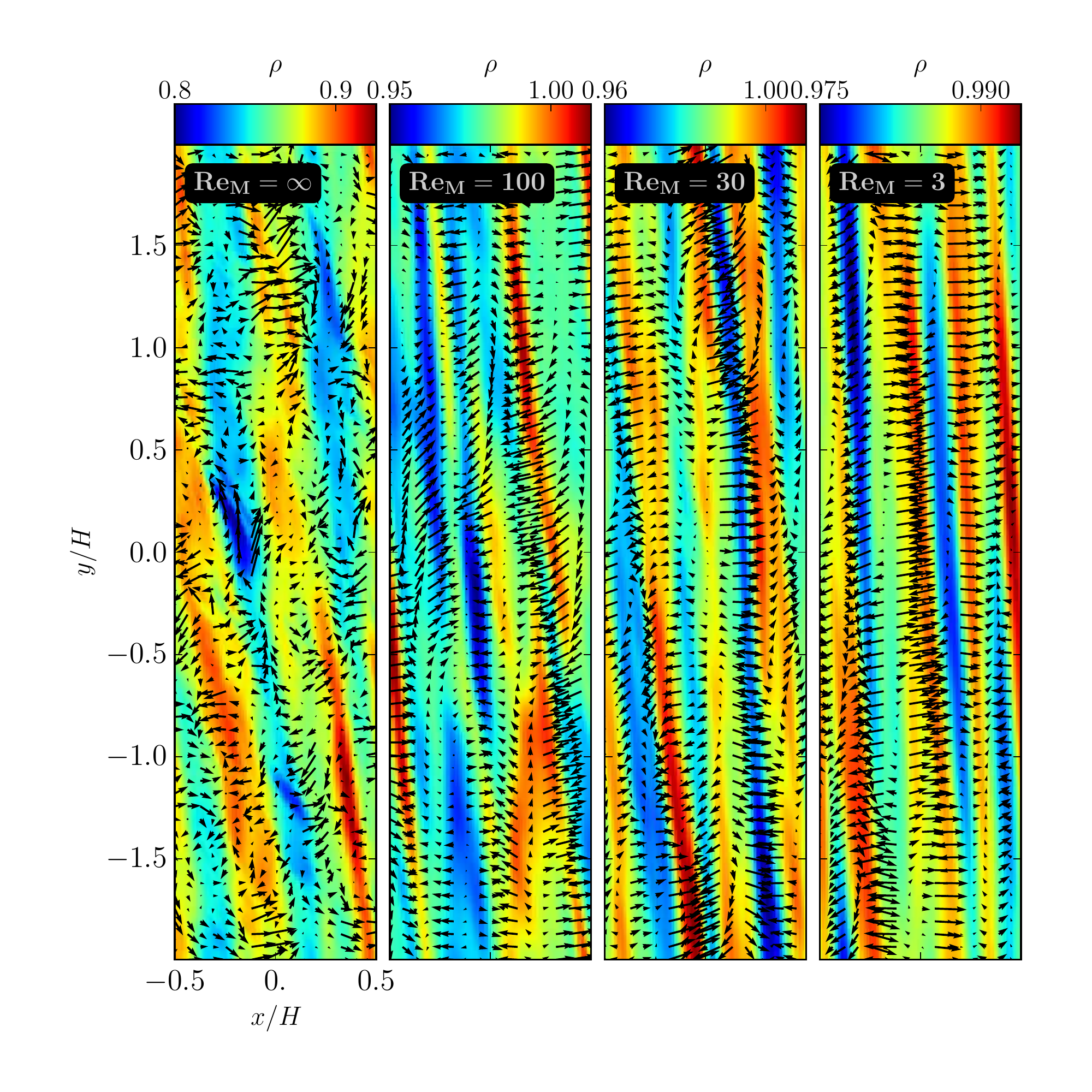}
 \caption{
    \label{f:img_64R_density_vel_xy}
    Slices through the midplane of density (color) with arrows
    following the in-plane ($x$-$y$) velocity.
  Each density image is scaled to its own
    minimum and maximum to emphasize morphology.}
\end{figure}

\begin{figure}
 \plotone{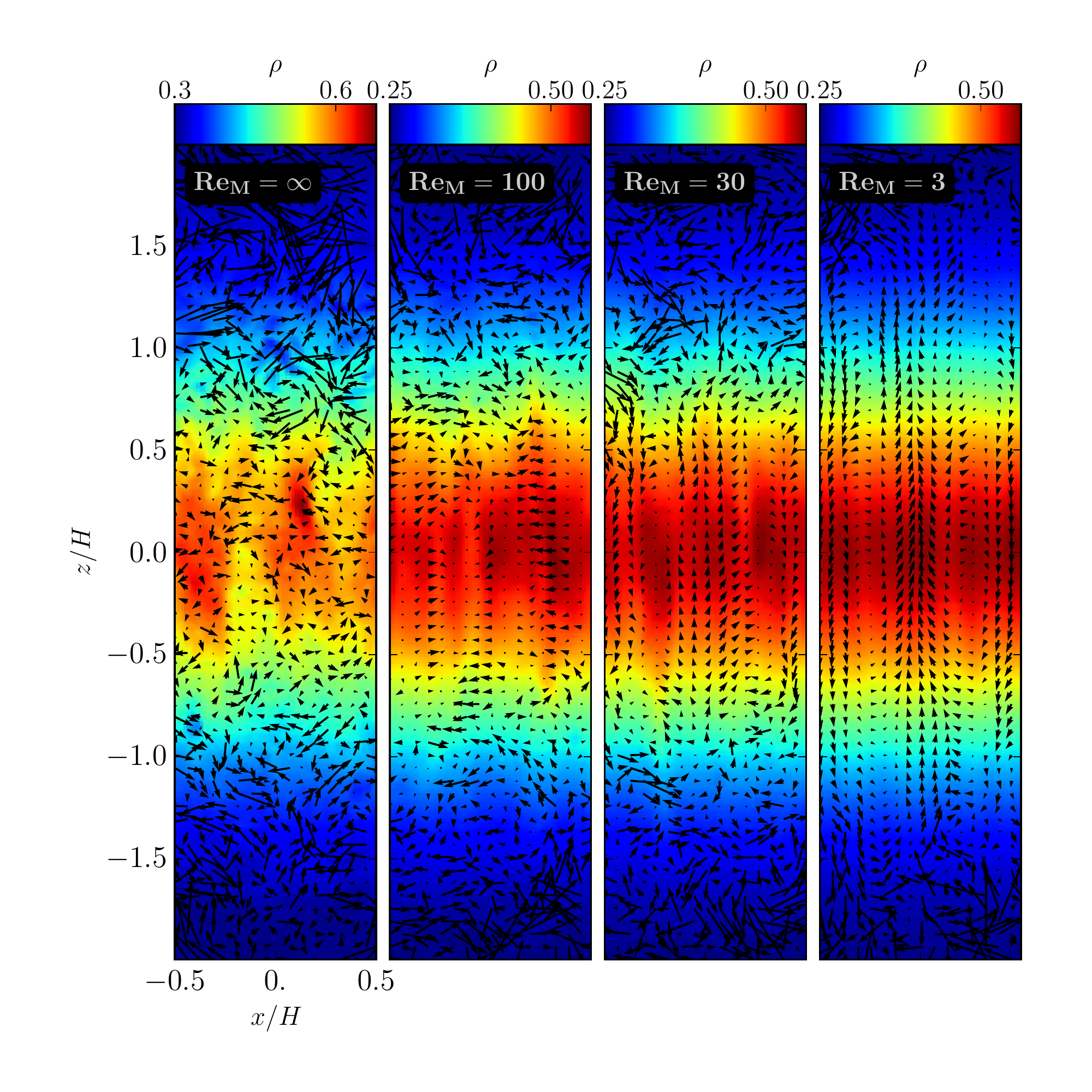}
 \caption{
    \label{f:img_64R_density_vel_xz}
    Slices through the $y=0$ (radial-vertical) plane of density (color) with arrows
    following the in-plane  ($x$-$z$) velocity.
  Each density image is scaled to its own
    minimum and maximum to emphasize morphology. Note the vertical
    circulation in dead zones viewed in this plane. }
\end{figure}

Figure~\ref{f:img_64R_density_vel_xy} shows images of density overlaid
with velocity vectors for the midplane (the $x$-$y$ plane at $z = 0$)
and Figure~\ref{f:img_64R_density_vel_xz} the $x$-$z$
plane at $y = 0$ for each of the FS runs. In the dead zone midplanes
(the right three panels), compressive waves dominate the velocity
field: velocity vectors lie almost entirely orthogonal to the density
striations in the $x$-$y$ plane. However, in the $x$-$z$ plane, inertial
mode signatures appear: the velocity is dominated by the $z$
component, traveling up and down in well defined vertical bands.

The dead zone shows ordered $u_z$ alternating between positive and
negative values across the dead zone that become more pronounced as
$Re_M$ decreases (Fig.~\ref{f:img_64R_density_vel_xz}, right three
panels). This suggests that the vertical oscillations are at $\theta =
0$, with energy transport from gravity waves purely vertical between
the active zones and the midplane. This is in stark contrast to the
fully MRI-active case. Near its midplane, there are no well-defined
vertical oscillations, and the $Re_M = 100$ model does not show nearly
as well-ordered motion as the larger dead zone models. This is perhaps
not surprising given that the fastest growing MRI mode has a growth
rate $q_{MRI} \sim \Omega_0$, which is greater than $N$ until around a
scale height. 

\subsection{Wave Modes}
The MRI is not operating in the densest parts of the disk in our dead
zone models, so we expect that the motions excited in this region will
take the form of linear wave modes stochastically excited by the
active zone turbulence. The following analysis is motivated by the
low-frequency oscillation in the volume-averaged kinetic energy most
clearly shown in the $Re_M = 3$ data between $ 60 < t < 100$ in
Figure~\ref{f:M64_energy_alpha_vorticity}. Fourier analysis of this
kinetic energy time series for runs with a dead zone show a clear peak
in frequency space that shifts slightly to lower frequency with
increasing dead zone size, suggesting the presence of a large
amplitude coherent oscillation in these runs.

In order to better understand these dead zone oscillations, we take
temporal power spectra of the radial and vertical velocity components
$u_x$, and $u_z$, and the density perturbation $\delta \rho =
\rho(x,y,z) - \rho_0(z)$. \citet{Arras+etal2006} and
\citet{Brandenburg2005} used similar methods to diagnose global modes
of oscillation in fully magnetorotationally turbulent disks. The
former were able to isolate acoustic and inertial
oscillations. However, the latter found no clear wave signatures for
MRI turbulence, though they did recover acoustic and inertial modes
for forced hydrodynamic turbulence in a shearing box.

\begin{figure}[htbp]
  \plotone{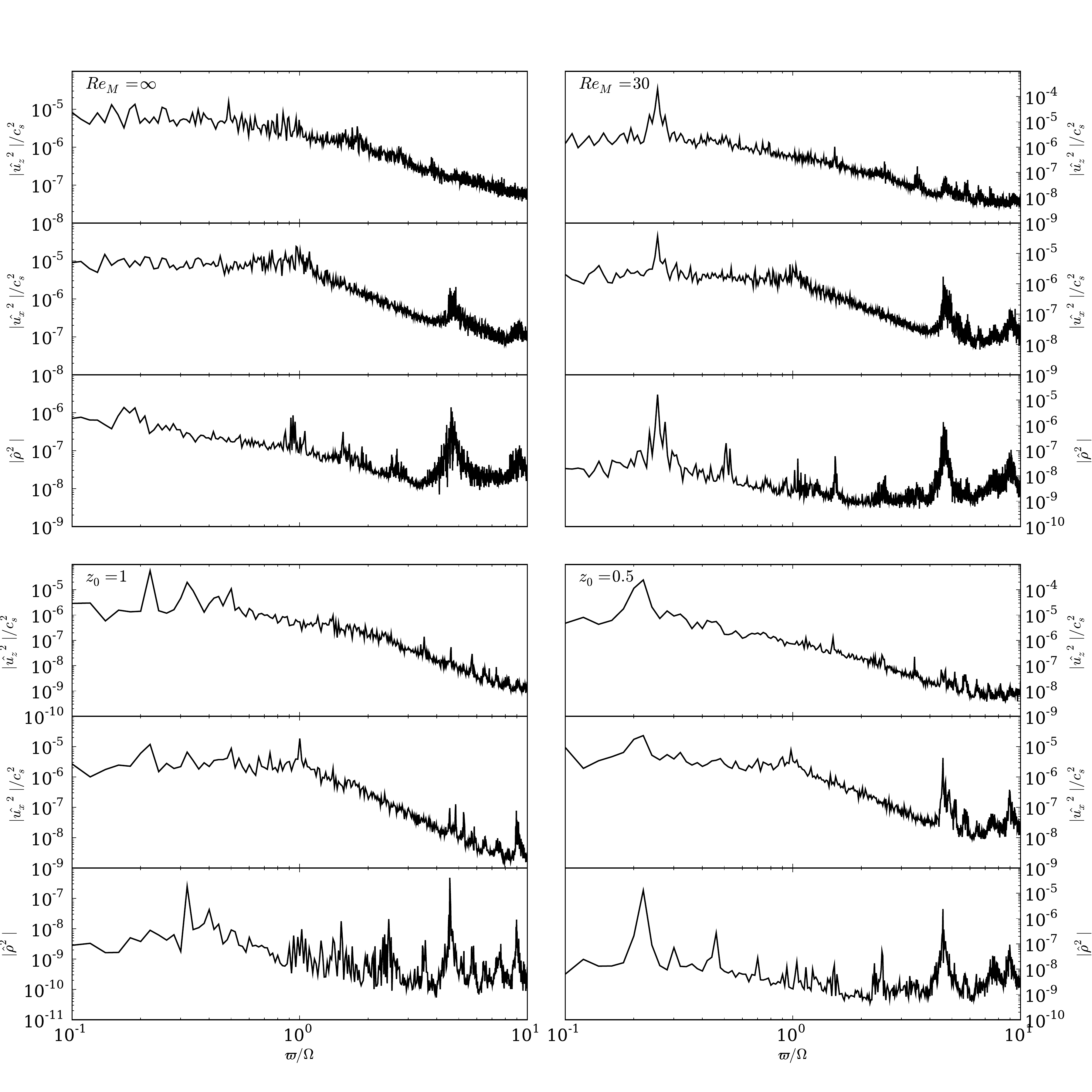}
  \caption{
    \label{f:yavg_t_spectra}
    Temporal power spectra for $u_x$, $u_y$, $\rho$. Acoustic modes
    are present in all runs at frequencies $> 1 \Omega$. Dead zone
    models also show a low-frequency inertial oscillation apparently
    driven by the active zones.  
  }
\end{figure}

Figure~\ref{f:yavg_t_spectra} shows spectra computed by taking an FFT
of a time series of each variable averaged in the $y$-direction at
every $(x,z)$ point on the grid of our low resolution model. The
resulting power spectra were then averaged to raise
signal-to-noise. Azimuthal averaging allows us to eliminate much of
the MRI power in the active zones \citep{Arras+etal2006}, though this
may not be a significant source in some of our larger dead zone
models.  The dominant peak in all variables at $\varpi_{max} \simeq
0.23 \Omega_0$ shifts only very slightly, to lower frequency, as the
dead zone is made thicker by a factor of two in the $\tanh$ runs. This
suggests that the vertical thickness of the dead zone is not setting
the oscillation frequency. We have also determined that this
characteristic frequency is unaffected by the radial ($x$) extent of
the box by running the $Re_M = 30$ model with $L_x = 2$ and confirming
that the oscillation frequency is unchanged.

Because the box is periodic, standing oscillations can be excited,
with discrete frequencies. The lowest frequency acoustic mode is in
the vertical direction and has a frequency $\varpi \simeq c_s 2
\pi/L_z = 1.11 \Omega_0$, which is clearly larger than
$\varpi_{max}$. The isothermal equation of state used in these models
precludes gravity waves (which would also have $\varpi < \Omega_0$),
and so we tentatively identify them as an inertial
oscillation. Furthermore, the small shift seen in the tanh runs as the
dead zone thickness increases is actually to \emph{lower} frequency,
while a higher boundary would drive internal oscillations at a
\emph{higher} frequency (see Eq.~\ref{e:brunt}) .

\subsubsection{Non-isothermal models and Buoyancy Forces}
\label{s:ideal_gas}
The low frequency of these oscillations is suggestive of a gravity
mode. However, these runs are strictly isothermal, $P = c_s^2 \rho$
with $c_s$ constant everywhere, precluding buoyant responses and thus
gravity waves. 

Recently, \citet{BaiGoodman2009} noted that isothermal equations of
state might spuriously enhance vertical mixing, as buoyant forces can
inhibit vertical mixing. Likewise, the lack of buoyancy in our
isothermal models may artificially enhance both the residual transport
at the midplane and the coherent oscillations. In order to understand
the effect of isothermality on these issues, we ran a single model at
$32$ zones per scale height including an ideal gas equation of state
$P = \rho k T/\mu$ and an entropy equation,
\begin{equation}
  \label{eq:entropy}
  \rho T(\frac{\partial s}{\partial t} + \mathbf{u \cdot \nabla} s +
  u_y^0 \frac{\partial s}{\partial y})
   =    \eta(z)\mu_0 \boldsymbol{j}^2
      + \zeta\rho\left(\boldsymbol{\nabla} \cdot u\right)^2
      + \kappa_6\boldsymbol{\nabla^6} s,
\end{equation}
where $\kappa_6$ is a hyperdiffusivity of entropy, $u_y^0 = -3/2
\Omega x$ is the Keplerian velocity profile, $\Omega$ is the rotation
rate of the shearing box, and all other symbols have their usual
meanings. The entropy equation includes resistive and shock heating
but not heating from the hyperdiffusivities (these terms are small,
and do not significantly contribute to the thermal balance).  We chose
a ratio of specific heats $\gamma = 5/3$. 

The boundary conditions on
this run as on the others are periodic in all directions, which
precludes the escape of heat.  We choose this rather unrealistic setup
to directly compare with our isothermal simulations, in order to
demonstrate that our results are not due to the lack of buoyancy in
the isothermal case. Because of the turbulent heating, the
total energy in the box increases compared to the isothermal
case. Although the turbulence is strongest in the surface layers,
there is no temperature inversion in the vertical direction: the
midplane remains hotter than the surface.
The disk quasi-statically adjusts to new equilibrium density and
pressure profiles as the energy increases.  While these profiles are
somewhat different from the standard Gaussian density profile expected
for accretion disks with linear $z$-gravity profiles,
Figure~\ref{f:en_adi_iso} shows that the main result of this
difference is a factor $\sim 2$ increase in magnetic energy, and a
factor $\sim 1.8$ increase in kinetic energy.

\begin{figure}
  \plotone{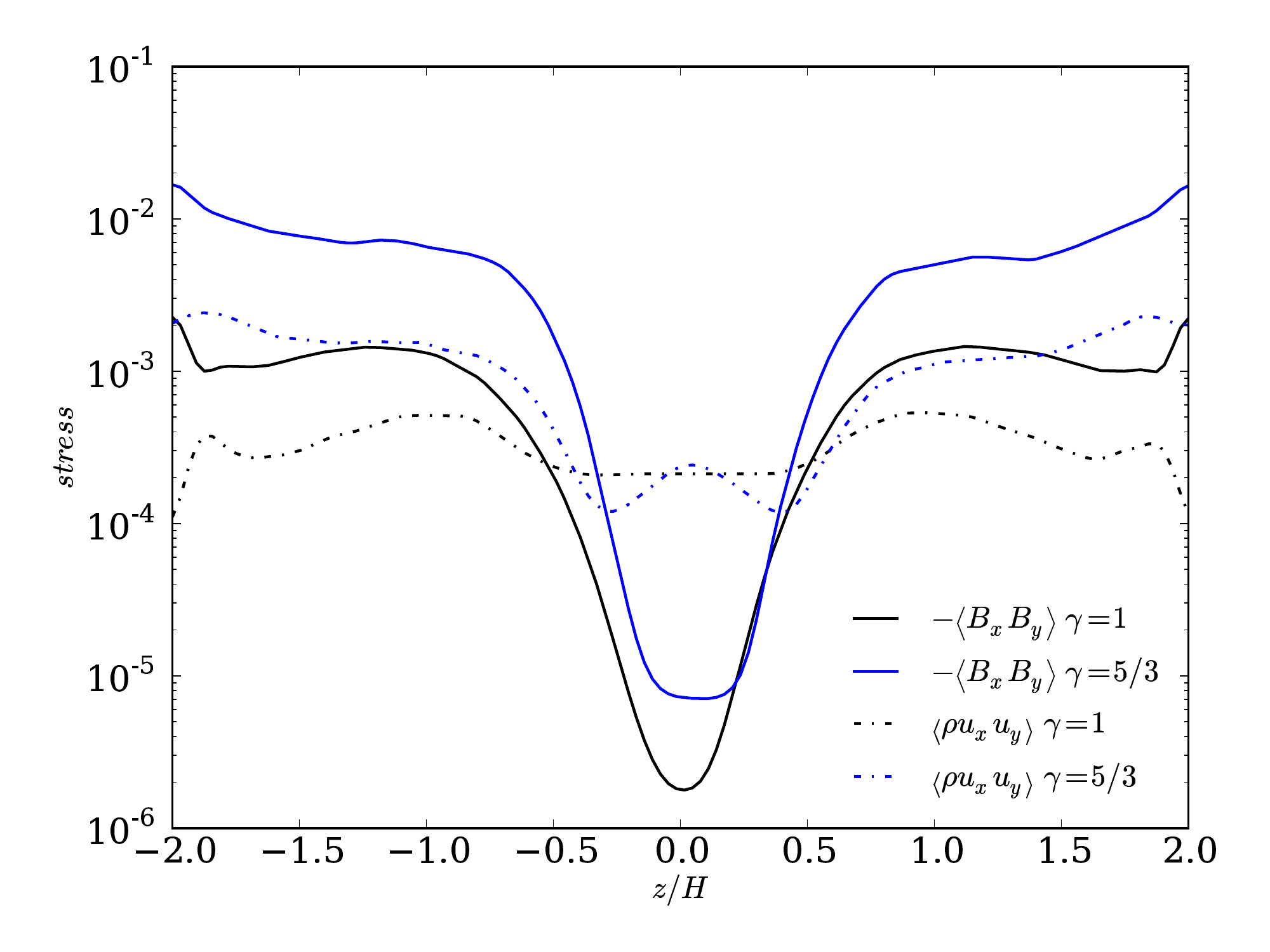}
  \caption{\label{f:stress_adi_iso}
 Stress vs z for isothermal and ideal gas runs.}
\end{figure}

\begin{figure}
  \includegraphics[height=8in]{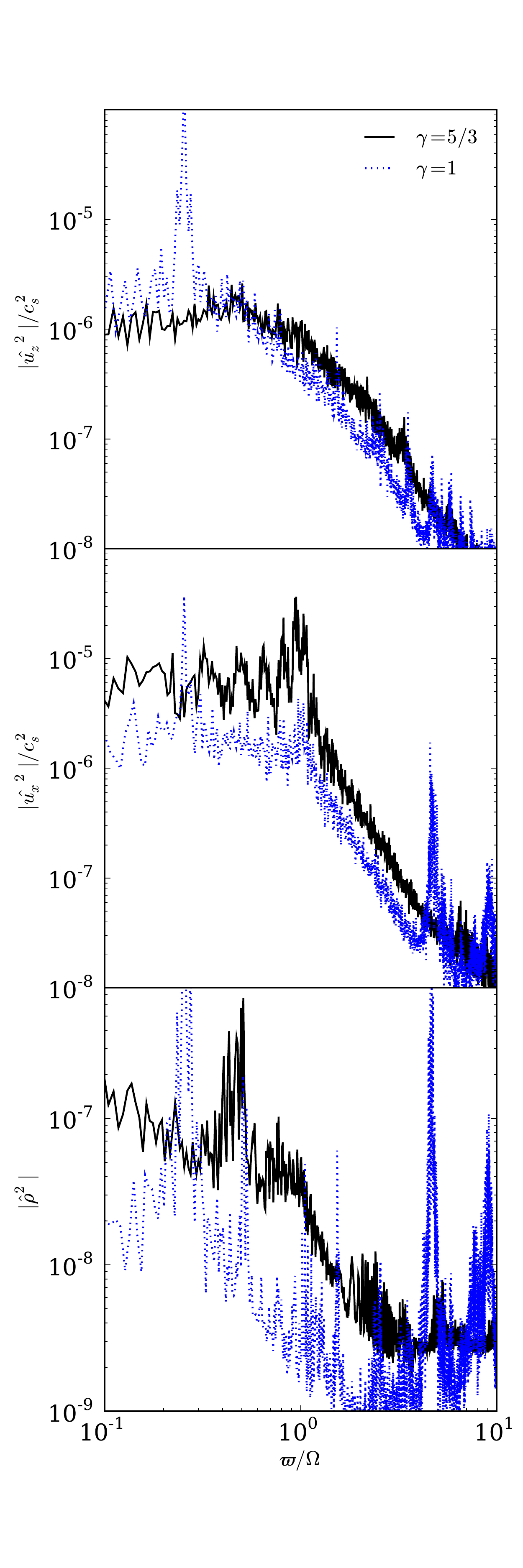}
  \caption{
    \label{f:adi_iso_spectra} Temporal power spectra of $u_x$, $u_y$,
    $\rho$ for $Re_M = 30$ with ideal gas equation of state (labeled
    $\gamma=5/3$) and isothermal ($\gamma = 1$) runs overplotted. Both
    inertial and acoustic waves are significantly damped but still
    present in the $\gamma=5/3$ case. The large amplitude mode at
    $\varpi \Omega \simeq 0.2$ in the isothermal case remains present,
    though at reduced amplitude and increased wavenumber $\varpi
    \Omega \simeq 0.5$. The normalization difference between the two
    spectra is due to the extra heat retained in the non-isothermal
    case.  }
\end{figure}

The overall morphology of the flow is roughly similar to the
isothermal model, with a reduced amplitude of motion in the dead zone
but a similar large-scale structure.  The stress profile across the
dead zone is narrower but deeper in the ideal gas run than the
isothermal one because of the pressure confinement caused by turbulent
heating in a periodic box absent cooling.  We directly compare the
Maxwell and Reynolds stresses for isothermal (labeled $\gamma = 1$)
and ideal gas ($\gamma=5/3$) in Figure~\ref{f:stress_adi_iso}.  We
find that inclusion of the buoyant forces reduces the amplitude of the
vertical motions and increases their characteristic frequency to
$\varpi_{max} \simeq 0.5 \Omega_0$, still well below the first
acoustic mode, as shown in Figure~\ref{f:adi_iso_spectra}. However,
the motions are not suppressed, even in this model in which turbulent
heating is not at all balanced by cooling.  A real disk in which
cooling is allowed will fall between the limits of the isothermal
model and this heating only model.  Therefore, we argue that the dead
zone motions represent a physical phenomenon even in the presence of
buoyancy forces.

\section{Midplane Vortex Dynamics}
\label{s:vortex}
We see no vortices in our simulations.  We must establish that this is
a physical result, and not due to insufficient resolution. JG05
demonstrate that sufficient numerical resolution is necessary to
sustain kinetic energy and angular momentum transport from
vortices. In their 2D simulations, they find $128 dx/H$ to be the
critical value, raising the question of whether vortices can form in
2D at our resolutions ($32 dx/H$ or $64 dx/H$)?  Furthermore, the
typical radial size scale of a vortex in their simulations is roughly
$H$, the entire width of our box. Can vortices form in such cramped quarters?

\subsection{Two-Dimensional Vortex Dynamics}
\begin{figure}
  \plotone{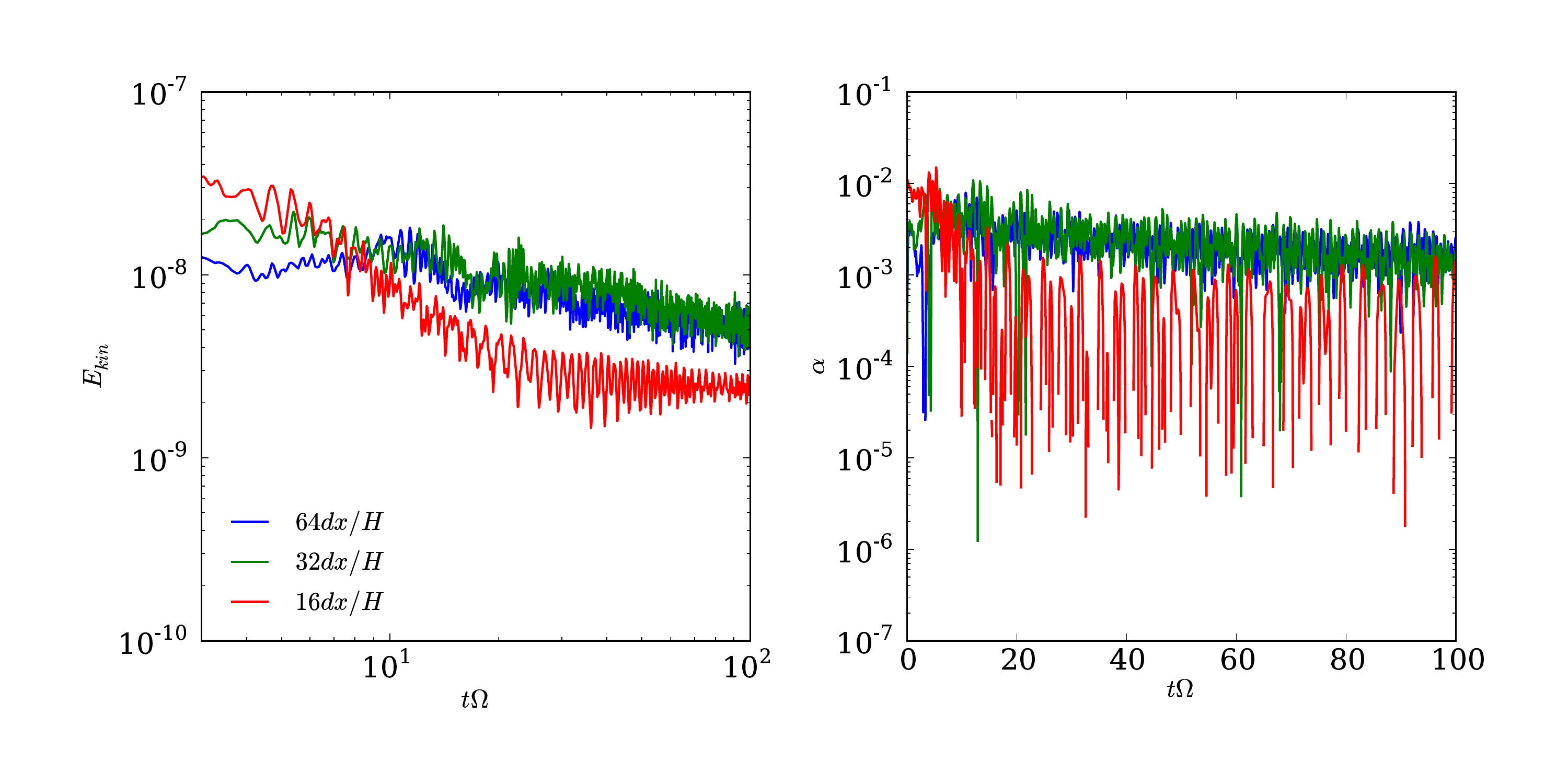}
  \caption{
    \label{f:JG_resolution}
    Kinetic energy and $\alpha$ for a 2D domain with $L_x = L_y = 4 H$ at three
    different resolutions, $(16, 32, 64) dx /H$, using initial
    velocity perturbations of magnitude $\sigma = 0.8 c_s$. Energy and
    $\alpha$ are clearly well converged at a resolution of $32 dx/H$.
  }
\end{figure}

In order to clarify these issues, we ran four sets of 2D hydrodynamic
simulations. First, we studied resolution.  A comparison between
Figure~\ref{f:incomp_shwave_t_res} and Figure~1 of JG05 shows that the sixth-order
Pencil Code is significantly less diffusive than their second-order, Zeus-derived
method, suggesting that we might be able to see sustained vortex
activity at lower resolution. Therefore we ran a set of models using
their domain with size $L_x = L_y = 4 H$ at resolutions of $(16, 32,
64) dx/H$, all with Gaussian random velocity perturbations with the
same initial perturbation amplitude, $\sigma = 0.8
c_s$. Figure~\ref{f:JG_resolution} shows the kinetic energy and
$\alpha$-parameter, $\alpha = \left< \Sigma u_x u_y \right>$ for 2D.
This Figure demonstrates clear convergence in these integrated
quantities at $32 dx/H$. 

\begin{figure}
  \includegraphics[scale=1.]{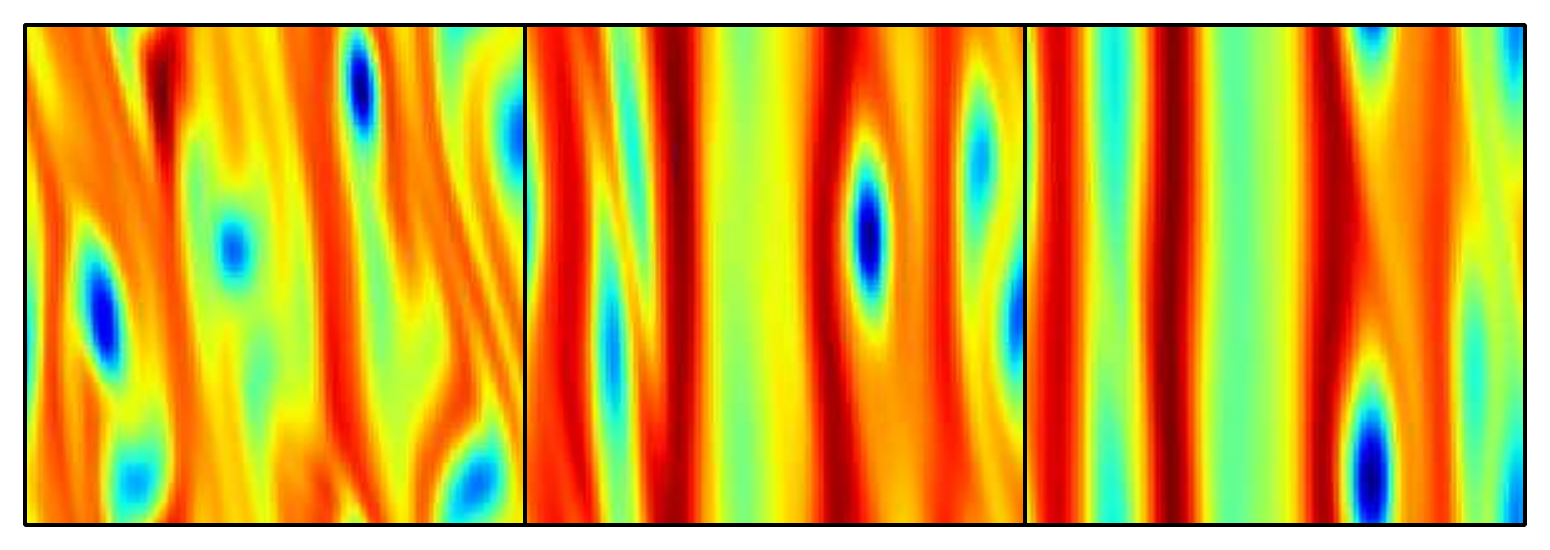}\\
  \includegraphics[scale=0.25]{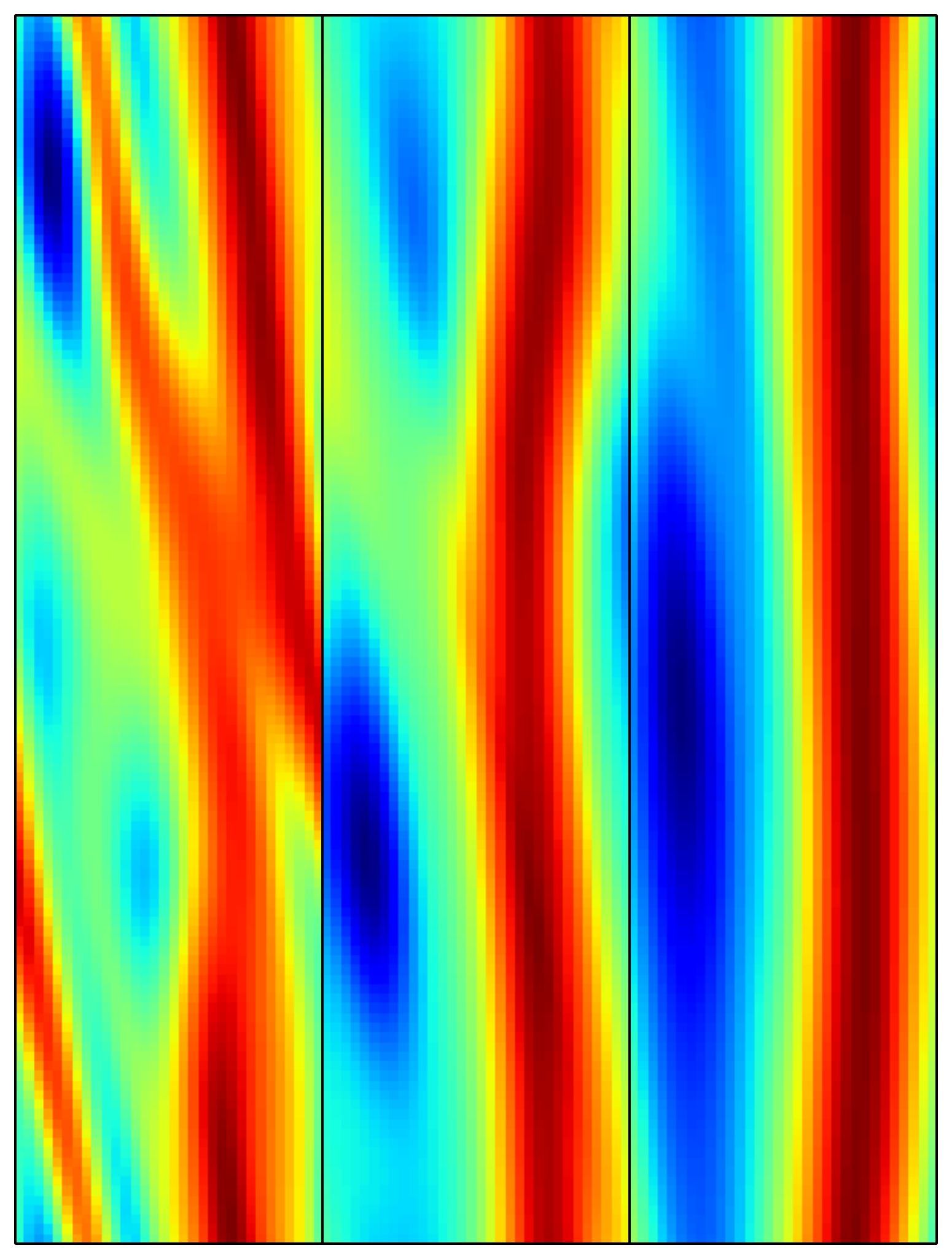}
  \caption{
    \label{f:JG_OMM_morphology} 
    Potential vorticity of the entire $x$-$y$ plane at times $t \Omega
    = 15.7, 31.4,47.1$ (left to right) \textit{upper panel} A $4H$
    square domain. Blue tones indicate negative potential vorticity,
    red positive. Vortex formation is similar to that found in
    \citet{JG05b}. \textit{lower panel} A domain $1 H \times 4 H$. In
    both panels, colors are scaled to the minimum and maximum of each
    image to highlight vortex morphology.  }
\end{figure}

Next, we turn to the question of domain size. JG05 use a domain four
times larger in radial extent than we do in our 3D simulations. Could
this inhibit vortex formation in our MHD runs? To test this, we ran a
set of 2D hydrodynamic models with the same domain as the midplane of
our 3D simulations, $L_x = 1 H$ and $L_y = 4 H$, again with
resolutions of $(16, 32, 64) dx/H$, seeded with perturbations of
magnitude $\sigma = 0.8 c_s$. Figure~\ref{f:JG_OMM_morphology} shows
that vortex morphology is largely similar in both domains at three
times, though there are more vortices present in the square domain. 

\begin{figure}
  \plotone{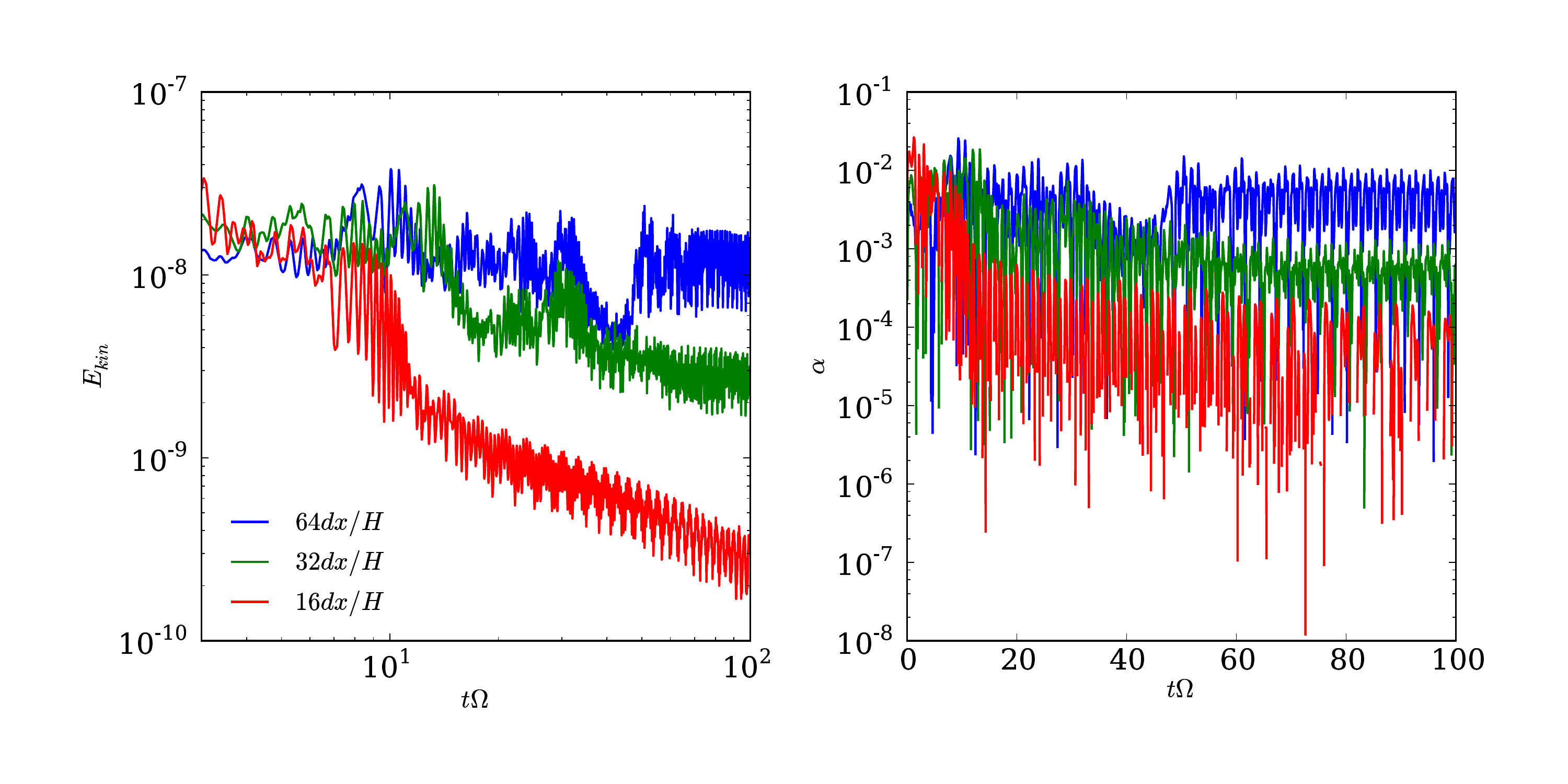}
  \caption{
    \label{f:OMM_resolution}
    Kinetic energy and $\alpha$ for a domain with $L_x = 1 H$, and
    $L_y = 4 H$, at three different resolutions, $(16, 32, 64) dx /H$,
    using initial velocity perturbations of magnitude $\sigma = 0.8
    c_s$.  Energy and $\alpha$ are not as well converged as in
    Figure~\ref{f:JG_resolution}, but sustained
    vortex activity does seem to occur even at $32\ \mathrm{zones/H}$.  }
  
\end{figure}

Figure~\ref{f:OMM_resolution} 
shows the resulting energy and $\alpha$. Although the results are not as well
converged as in the square domain, there is still evidence for
vortex activity with lifetimes clearly greater than the MRI growth
time $\tau_{\rm MRI}\sim \Omega$ at a resolution of $32 dx/\mathrm{H}$. 

\begin{figure}
  \plotone{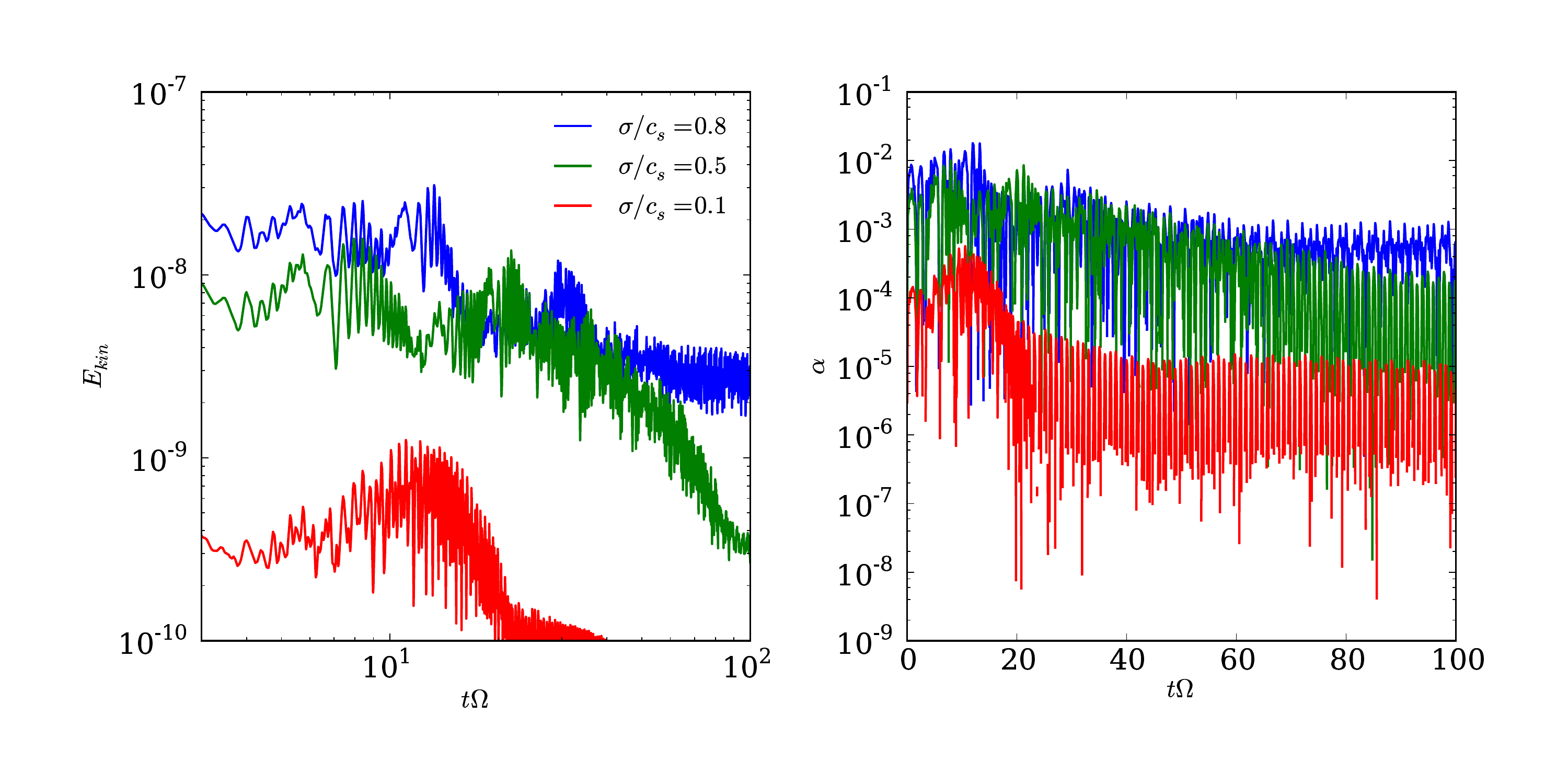}
  \caption{
    \label{f:OMM_intensity}
    Kinetic energy and $\alpha$ for a domain with $L_x = 1 H$ and $L_y
    = 4 H$ at resolution of $32 dx / H$, using three different initial
    velocity dispersions, $\sigma = (0.1, 0.5, 0.8) c_s$. Sustained
    vortex activity clearly requires strong initial perturbations.  }
\end{figure}

What, then, is the reason we do not see vortex activity in the 3D dead
zone simulations? We have argued that dimensionality is not the limiting factor,
as regions with lower Froude number that are more effectively 2D do
not show more vorticity. It also appears clear that the 2D vortex
activity found in previous simulations can be reproduced on domains
like ours at the resolutions we use in our MHD runs, so resolution and
grid size are also not limiting. Rather, the
discriminant appears simply to be the \emph{strength} of perturbations
required to trigger long-lived vortices. We ran a third set of 2D
hydrodynamic simulations, this time at a fixed resolution of $32 dx/
\mathrm{/H}$, but with decreasing velocity perturbation magnitudes $\sigma=
0.8 c_s, 0.5 c_s$, and $0.1
c_s$. Figure~\ref{f:OMM_intensity} shows that below $\sigma = 0.8$,
the kinetic energy and $\alpha$ drop precipitously after only a few
orbits. (Note that we keep the domain size fixed at $L_x = 1 H$, $L_y =
4 H$, unlike in the tests of JG05, in which the domain size was scaled linearly with
the strength of the velocity dispersion.) We conclude 
that \emph{compressible} vortices require strong $\sigma =
0.8 c_s$ initial perturbations in order to survive for many orbits.

\citet{UmurhanRegev2004} see perturbation energy sustained essentially
indefinitely for $Re = \infty$ and decaying away only very slowly for
$Re = 50000$, but their model is incompressible. They show this
approximation to be rigorously valid for the small length scales they
consider. However, interestingly, they find that the perturbation energy saturates at
roughly $\lesssim 10^{-2} E_{shear}$, where $E_{shear}$ is the energy in
the shear. They define the turbulent intensity of an incompressible
shear flow as
\begin{equation}
  \varepsilon_0 = \frac{\int u^2 \, \ud x \,  \ud y}{\int u_{sh}^2 \, \ud x \, \ud y} \lesssim 10^{-2},
\end{equation}
with $u$ the disturbance velocity and $u_{sh}$ the shear velocity. We
write $u_{sh} = q \Omega x$, integrate for the Keplerian case $q =
-3/2$, and use our definition of (thermal) scale height $H =
2 c_s/\Omega$ to arrive at
\begin{equation}
  \langle \mbox{Ma}^2 \rangle = \frac{\sigma^2}{c_s^2} = \frac38 \varepsilon_0 \left(\frac{L_x}{H}\right)^{2},
\end{equation}
with $\langle \mbox{Ma}^2 \rangle^{0.5} = \mbox{Ma}_{rms}$ the RMS
Mach number.  Using their saturation value, this gives
$\mbox{Ma}_{rms} \sim 0.06 (L_x/H)$, or about $0.06$ over a domain of
$H$. This is an order of magnitude below the levels necessary to
incite vortices on the larger scales that we simulate in which
compression becomes important. If we assume that small scale,
non-linear, transient growth instability scales with Reynolds number
as found by \citet{LesurLongaretti2005}, not only is the
incompressible, small-scale, transient growth irrelevant for angular
momentum transport, its saturation level is so small that larger scale
vortices in the compressible regime will not form at all.

\subsection{Long-lived axisymmetric structures}
Recently, \citet{JohansenYoudinKlahr2009} reported the existence of
long-lived axisymmetric zonal flows in MRI turbulence, provided the
radial width of the box is larger than $\sim 1 H$. These structures
result from an inverse cascade of magnetic energy to the largest
scales present in the box. This energy drives a large scale Maxwell
stress, which in concert with the Coriolis force enforces the zonal
flow. The end result are long-lived ($\sim 10$s of orbits) pressure
and density fluctuations. It is easy to imagine that these complex
structures could have significant influence on motions in the dead
zones, assuming there is sufficient magnetic energy to launch the
zonal flows in the first place. Figure~\ref{f:maxw_z_32} implies that
this criterion is unlikely to be met: even in the moderate sized dead
zone models (e.g., $Re_M = 30$ and $z_0 = 1$), the total Maxwell
stress is many orders of magnitude lower than that in the MRI active
zones. However, given that \citet{JohansenYoudinKlahr2009} demonstrate
that the zonal flows are subtle, stochastically driven, and
non-linear, we find such a crude estimate to be unsatisfactory.

\begin{figure}
  \includegraphics[scale=0.5]{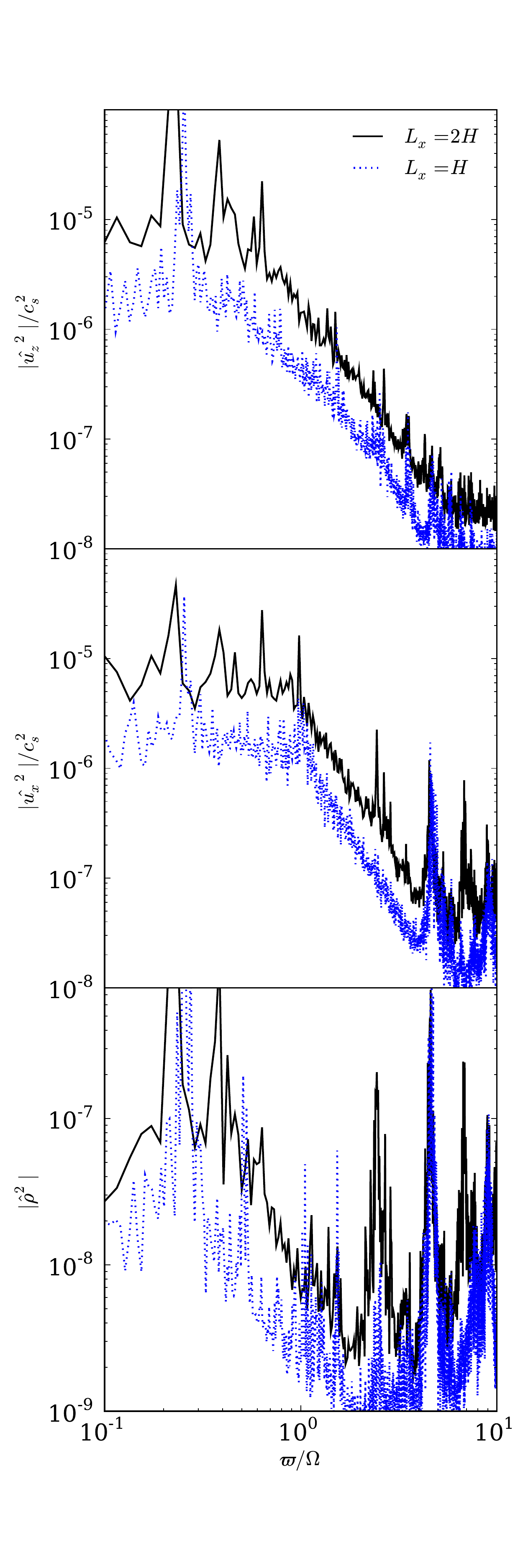}
  \caption{\label{f:yavg_t_spectra_x2h} Temporal power spectra of
    $u_x$, $u_y$, $\rho$ for $Re_M = 30$ with the standard box width
    ($-0.5 H < x < 0.5 H$) and a box twice as wide ($-H < x < H$). The
    wider box features the large axisymmetric pressure bumps reported
    by \citet{JohansenYoudinKlahr2009}, but this has little effect on
    the oscillations reported in this paper.
  }
\end{figure}

Therefore, we re-ran our $Re_M = 30$ FS model with a radially-enlarged
domain, running from $ -H < x < H$. By averaging $\rho$ over the
entire $y-z$ domain, including both active and dead zones, we recover
the reported long-lived axisymmetric structures. Nevertheless, our
results remain robust: there is no vortex formation, and
figure~\ref{f:yavg_t_spectra_x2h} demonstrates that the midplane
oscillations are quite similar between the larger domain and our
fiducial case.

\section{Discussion}
\label{s:discussion}
We find a complete lack of coherent anticyclonic vortices in all of
our stratified local simulations of dead zones, despite the presence
of a vorticity source in the form of active zones. While our
experiments do not tell us exactly what happens, it appears that even
if the midplane were a pure 2D layer, large scale vortex formation
would not occur because the amplitude of velocity perturbations driven
in the dead zone by the active layers is well below that
needed. Recent work by \citet{LesurPapaloizou2009} suggests that
vortices in 3D disks are always parametrically unstable, though the
growth rates of such instabilities can be very slow. Even in the
presence of an active vorticity source, though, we find no vortex
formation.

Our simulations cover a smaller range of $z$ than
\citet{BarrancoMarcus2005}, and it is possible that the increased
stratification present at higher $z$ may change our results. To
strengthen the results presented here, direct comparison with
\citet{BarrancoMarcus2005} including compressibility would be quite
enlightening. Furthermore, almost all of our simulations use an
isothermal relation for pressure and density. This forces density
gradients and pressure gradients to coincide, eliminating the
baroclinic term driving the vortex formation from breaking gravity
waves in \citet{BarrancoMarcus2005}. Relaxing this requirement in our
simulations may aid vortex formation at higher scale heights though
our one adiabatic model without cooling did not show vortex formation
either.

However, we do find a large-scale, axisymmetric oscillation about the
midplane in all dead zone models. This oscillation carries nearly all
the kinetic energy of the dead zone. We are unable to associate it
with a normal mode using a simple linear dispersion relation. A more
detailed analysis including the effects of continuous stratification
could shed considerable light on the situation. Regardless of such
details, the low frequency (less than $\Omega_0$) suggests that the
mode will not have much effect on dust sedimentation, which is
typically dominated by random fluctuation with correlation times
$\tau_{corr} \simeq 0.15 t_{orb}$, corresponding to frequencies
$\simeq 7 \Omega_0$ \citep{FromangPapaloizou2006}. It could, in
principle, cause a warp in the dust layer, which in turn could affect
the gravitational stability
\citep{GoldreichWard1973,Johansen+etal2007}.  Such phenomena can only
be examined with a combined treatment of dust, self-gravity, and a
dead zone, which we defer to future work.

Finally, global dynamics may significantly alter vortex formation
properties. A real protoplanetary disk is thin and quasi-2D for
horizontal scales greater than the scale height. It is quite possible
that robust, anticyclonic vortices may form within a disk, but our
results suggest that a local mechanism is not responsible. The
baroclinic instability
\citep{KlahrBodenheimer2003,PetersenJulienStewart2007,PetersenStewartJulien2007}
remains a candidate for forming such large scale vortices.

\acknowledgements
JSO would like to thank the Hausdorff Research Institute for
Mathematics at Universt\"at Bonn, where a considerable portion of this
work was completed, for their hospitality and financial
support. J. Hawley, K. Menou, and P. Arras made useful suggestions on
an early version of this manuscript. Computations were performed at
the Parallel Computing Facility of AMNH and the Cray XT3 ``Big Ben''
at the Pittsburgh Supercomputing Center, the latter of which is
supported by the National Science Foundation.  M-MML was partially
supported by NASA Origins of Solar Systems grant NNX07AI74G and by NSF
grant AST08-35734.

\clearpage

\clearpage

\clearpage

\clearpage

\clearpage

\clearpage

\clearpage

\clearpage

\clearpage

\clearpage

\clearpage

\clearpage

\clearpage

\clearpage

\clearpage

\clearpage

\clearpage

\clearpage

\clearpage

\clearpage

\begin{deluxetable}{lllllllll}

\tablecaption{
\label{t:enstrophy}
Directional normalized enstrophy for dead and active zones.
}

\tablehead{& \multicolumn{3}{c}{Active} & \multicolumn{3}{c}{Dead}\\
 \colhead{Run} & \colhead{$\omega_x^2/\omega^2$} &
 \colhead{$\omega_y^2/\omega^2$} & \colhead{$\omega_z^2/\omega^2$} &
 \colhead{$\omega_x^2/\omega^2$} & \colhead{$\omega_y^2/\omega^2$} &
 \colhead{$\omega_z^2/\omega^2$}
}
\startdata
64Rinf & 0.31 & 0.34 & 0.35 & \nodata & \nodata  & \nodata \\
64R100 & 0.32 & 0.34 & 0.35 & 0.11 & 0.69 & 0.20\\
64R30  & 0.35 & 0.36 & 0.29 & 0.07 & 0.79 & 0.14\\
64R3   & 0.31 & 0.41 & 0.28 & 0.06 & 0.68 & 0.26\\
32z0.5 & 0.26 & 0.48 & 0.27 & 0.10 & 0.68 & 0.22\\
32z1.0 & 0.27 & 0.46 & 0.26 & 0.07 & 0.76 & 0.17\\
\enddata
\end{deluxetable}

\end{document}